\documentclass[11pt,notitlepage,tightenlines,eqsecnum,floats,aps,amsmath,superscriptaddress,amssymb,nofootinbib,longbibliography]{revtex4-1}

\usepackage{graphicx, wrapfig}
\usepackage{amssymb}
\usepackage[usenames, dvipsnames]{color}
\usepackage{diagbox}
\usepackage{mathrsfs}
\usepackage{subfigure}
\usepackage{float}
\usepackage{slashbox}
\usepackage{verbatim}
\usepackage{cancel}
\usepackage[normalem]{ulem}
\usepackage[hidelinks]{hyperref}
\hypersetup{
  colorlinks   = true, %Colours links instead of ugly boxes
  urlcolor     = blue, %Colour for external hyperlinks
  linkcolor    = blue, %Colour of internal links
  citecolor   = blue %Colour of citations
}
\usepackage{enumerate}

\newcommand{\be}{\begin{equation}}
\newcommand{\ee}{\end{equation}}
\newcommand{\bea}{\begin{eqnarray}}
\newcommand{\eea}{\end{eqnarray}}

\def\f{\frac}

\def\dd{{\rm d}}
\def\pp{p_{\phi}}
\def\f{\frac}
\def\h{\hat}
\def\l{\left}
\def\r{\right}
\def\dd{\textrm{d}}
\def\d{\textrm{d}}

\begin{document}

\title{Observational constraints on anisotropies for bouncing alternatives to inflation}

\author{Ivan Agullo}\email{agullo@lsu.edu}
\affiliation{Department of Physics and Astronomy,
Louisiana State University, Baton Rouge, Louisiana 70803, USA}

\author{Javier Olmedo}\email{javolmedo@ugr.es}
\affiliation{Departamento de F\'isica Te\'orica y del Cosmos,
Universidad de Granada, Granada-18071, Spain}

\author{Edward Wilson-Ewing} \email{edward.wilson-ewing@unb.ca}
\affiliation{Department of Mathematics and Statistics,
University of New Brunswick, Fredericton, NB, Canada E3B 5A3}

\begin{abstract}

We calculate how primordial anisotropies in the background space-time affect the evolution of cosmological perturbations for bouncing alternatives to inflation, like ekpyrosis and the matter bounce scenario. We find that the leading order effect of anisotropies in the contracting phase of the universe is to induce anisotropies in the cosmic microwave background with a very concrete form: a scale-invariant quadrupolar angular distribution. Sub-leading effects are the generation of higher-order moments in the angular distribution, as well as cross-correlations between scalar and tensor modes. We also find that observational constraints from the cosmic microwave background on the quadrupole moment provide strong bounds on allowed anisotropies for bouncing alternatives to inflation that are significantly more constraining than the bounds previously obtained using scaling arguments based on the conjectured Belinski-Khalatnikov-Lifshitz instability.

\end{abstract}

\maketitle

\section{Introduction}
\label{intro}

Cosmological models where the big bang singularity is replaced by a cosmic bounce have emerged as interesting alternatives to inflation for explaining the origin of the primordial density perturbations observed in the cosmic microwave background (CMB). But a bounce on its own is not sufficient: in the most popular proposals, the primordial perturbations are generated during the contracting phase by a mechanism unrelated to the bounce, and which varies among models. Perhaps the two most studied cosmologies of this type are ekpyrosis and the matter bounce scenario. In ekpyrosis, the contraction is dominated by a form of matter with pressure greater than the energy density, giving a nearly scale-invariant spectrum of entropy perturbations which can, in turn, source nearly scale-invariant adiabatic perturbations \cite{Khoury:2001wf, Lehners:2008vx}. For the matter bounce, the scale-invariant primordial power spectrum is generated during a phase of pressureless matter-dominated contraction \cite{Wands:1998yp, Finelli:2001sr}.

Despite the healthy competition these models provide to the inflationary paradigm, a challenge all bouncing models must face is the instability to the growth of anisotropies during the contracting pre-bounce phase. During contraction, the contribution from anisotropic stresses to the Friedman equation grows very rapidly, much faster than the energy density of common matter fields like radiation, baryonic matter, and cold dark matter, and will eventually come to dominate the dynamics in a contracting cosmology over these types of matter fields. Quantitatively, we can typically expect the ratio of the anisotropic stresses $\sigma^2$ to the total energy density $\rho$ to eventually become greater than unity in a contracting space-time, $\sigma^2 / (16 \pi\, G\, \rho) \gtrsim 1$. Once this ratio becomes comparable to 1, it is not possible to assume the background space-time is (approximately) isotropic. While isotropy is not required for a perturbative treatment of inhomogeneities, most of the standard tools of cosmological perturbation theory assume an isotropic background which greatly simplifies the analysis (see, e.g., \cite{Mukhanov:1990me}) compared to an anisotropic background geometry.

In presence of inhomogeneities, the situation is even more difficult---in this case, the ratio $\sigma^2 / (16 \pi\, G\, \rho)$ (which now will depend on position as well as time) becoming larger than 1 is expected to indicate the onset of the conjectured Belinski-Khalatnikov-Lifshitz (BKL) chaotic instability \cite{Belinsky:1970ew} and the associated loss of predictivity due to the high sensitivity of the dynamics to the initial conditions, with neighbouring points eventually following very different dynamics. If this does occur, the large anisotropies will impact upon the density fluctuations in a highly inhomogeneous manner and in such a case it seems difficult to produce the nearly isotropic temperature fluctuations observed in the cosmic microwave background (unless nearly scale-invariant perturbations can be generated after the bounce, once anisotropies are small, as may be possible for example in bouncing inflation models).

Different strategies have been proposed to avoid the conjectured BKL instability. The simplest is to assume that anisotropies were initially zero, or extraordinarily small such that they always remain subdominant up to and including the bounce, but this obviously introduces a fine-tuning problem \cite{Bozza:2009jx, Levy:2016xcl}. A less drastic assumption is to assume that the last stages of contraction included an ekpyrotic phase \cite{Khoury:2001wf, Qiu:2013eoa, Cai:2013kja}, in which the universe is dominated by ``ultra-stiff'' matter with a pressure $p$ that is greater than its energy density $\rho$, so the equation of state $w$ satisfies $w = p / \rho > 1$. For such ekpyrotic matter, in a contracting universe $\rho$ grows faster than the anisotropic stress $\sigma^2$, so the ratio $\sigma^2/(16 \pi\, G\, \rho)$ decreases as the universe contracts, and it is possible for anisotropies to always remain subdominant without requiring significant fine-tuning, although it is still necessary to choose initial conditions so the ekpyrotic field starts to dominate the dynamics of the contracting cosmology before anisotropies do (note that an instability to other sources of anisotropies could still survive in ekpyrotic models \cite{Barrow:2015wfa}). Finally, modified gravity models have also been proposed as a possible solution to the unstable growth of anisotropies in bouncing scenarios \cite{Lin:2017fec}.

In addition to avoiding the BKL instability, there is another good reason to require that anisotropies always remain small: observational constraints. In this paper, we show that the CMB provides strong constraints on anisotropies in bouncing cosmological models. Since these constraints come from observational data, they are more direct than order of magnitude arguments based on the BKL conjecture. More importantly, we show that the CMB data provide stronger bounds on primordial anisotropies in bouncing alternatives to inflation, as compared to simply requiring that the ratio $\sigma^2/(16 \pi\, G\, \rho)$ must always remain smaller than 1 \cite{Cai:2013vm, Xue:2013bva}. Finally, these observational constraints are also relevant for ekpyrotic cosmology, which is weakly constrained by the condition $\sigma^2/(16 \pi\, G\, \rho) \ll 1$.

Cosmological perturbations are a powerful probe of anisotropies in the early universe. The reason why CMB constraints are stronger is because their anisotropic features do not necessarily dilute with the expansion.
This is in sharp contrast to the anisotropies in the homogeneous background geometry that are rapidly diluted in an expanding universe (even if the anisotropies are initially large). Since perturbations retain some memory of any early anisotropic phase and imprint this in the CMB, the extraordinary isotropy observed in the CMB provides stringent constraints on anisotropies in bouncing cosmologies.

The main challenge of our analysis is to compute the evolution of cosmological perturbations in anisotropic space-times, a task significantly more tedious than its counterpart in isotropic scenarios. We overcome this difficulty by taking advantage of recent developments in the classical \cite{Pereira:2007yy, Pitrou:2008gk, Agullo:2020uii, Agullo:2020kil} and quantum theory \cite{Agullo:2020uii} of cosmological perturbations in anisotropic Bianchi~I space-times. (For earlier work on cosmological perturbations with an anisotropic background, see \cite{Doroshkevich:1971, Perko:1972cs, Tomita:1985me, Noh:1987vk, Dunsby:1993fg}, and for the calculation of the power spectrum for a test scalar field on an anisotropic contracting background, see \cite{Modan:2022fzg}.) The Bianchi~I space-time is a generalization of spatially flat Friedman-Lema\^itre-Robertson-Walker (FLRW) metrics where the scale factors in the three Cartesian directions of space evolve independently; this is the simplest of the family of anisotropic but homogeneous metrics. As could be expected, anisotropies in the background homogeneous space-time will source anisotropic features in the perturbations, but the exact way this occurs is rather complicated: it is not only the ratio $\sigma^2/(16\pi\, G\, \rho)$ that is important, but other powers $\sigma^n/\rho^m$ also become relevant for the evolution of perturbations. Furthermore, perturbations acquire anisotropic features over time in a cumulative manner; consequently, the final power spectrum can contain prohibitively large anisotropies even if the ratio $\sigma^2/(16\pi\, G\, \rho)$ is significantly smaller than unity at all times during the past history of the universe.

A further goal of this paper is to compute the concrete form of the anisotropic imprints expected in the CMB from bouncing cosmologies. In particular, we find that under mild assumptions, a model-independent prediction of bouncing alternatives to inflation is a nearly {\em scale-invariant} spectrum of anisotropies, with an angular dependence dominated by a {\em quadrupole}---this appears to be a generic fingerprint of these models of the early universe. Further, the amplitude of the quadrupole has exactly the same red tilt as the scalar power spectrum. We also find that anisotropies introduce cross-correlations between scalar and tensor modes, as has already been pointed out \cite{Doroshkevich:1971, Perko:1972cs, Tomita:1985me}; the importance of this second effect depends on the initial amplitude of the tensor modes and therefore is model-dependent.

Note that any cosmological model with a bounce requires some form of new physics, whether matter fields that violate energy conditions or modifications to general relativity, perhaps motivated by quantum gravity. The main results derived here do not significantly depend on the physics of the bounce. Recall that in both ekpyrosis and the matter bounce, nearly scale-invariant perturbations are generated at a time well before the bounce and the new physics that causes the bounce can safely be neglected during the contraction phase when the near scale-invariance is generated. Similarly, the largest part of the anisotropic features imprinted onto the linear perturbations come from the contracting phase, and it is independent of the new physics ingredient that causes the bounce. For concreteness, in the simulations presented in this paper we assume the bounce follows the dynamics predicted by loop quantum cosmology, but the main constraints are independent of this choice. 

The rest of this article is organized as follows. We begin in Sec.~\ref{constCMB} with a summary of the constraints on anisotropies from CMB observations that are most relevant for this work. In Sec.~\ref{dyn}, we describe the dynamics of both the homogeneous background and perturbations, and explore the evolution of anisotropies with the aid of numerical simulations. For the background, we consider a contracting Bianchi~I universe dominated by three different matter contents, radiation fluid ($w=1/3$), a stiff fluid ($w=1$) and an ekpyrotic fluid (with $w=3$), in order to study how the dynamics vary depending on the matter content of the universe. In Sec.~\ref{upp} we compare our calculations with the CMB data summarized in Sec.~\ref{constCMB}, and extract upper bounds for the anisotropic shear for each of the three types of matter content considered. Finally, in Sec.~\ref{pred} we compute the form of the primordial power spectrum and angular correlation functions in the CMB expected from an anisotropic bounce. Throughout the paper we work in Planck units where $\hbar=c=G=1$.

\section{Constraints from the CMB}
\label{constCMB}

This section provides a brief summary of constraints from the Planck Collaboration on anisotropic features in the CMB \cite{Planck:2019evm, Planck:2018jri}; these constraints will guide the analysis we do in this paper. (See also \cite{Collins:1973lda, Maartens:1994qq, Martinez-Gonzalez:1995, Kogut:1997az, Stoeger:1997puj, Kim:2013gka, Ramazanov:2013wea, Rubtsov:2014yua, Saadeh:2016sak} for earlier constraints on anisotropic features in the CMB.)

One of the goals of the Planck satellite was to test the hypothesis of homogeneity and isotropy on which the standard cosmological model rests. As usual, in order to extract quantitative conclusions from data, one first needs to introduce a mathematical model for the features one is looking for, containing some free parameters to be constrained using observations. Here we are interested in constraints on anisotropies (i.e., preferred directions) in the CMB, potentially sourced by an anisotropic background corresponding to a Bianchi space-time.

One way of parameterizing primordial anisotropies is by allowing the scalar power spectrum of comoving curvature perturbations, ${\cal P}_{\mathcal{R}}(\vec k)$, to depend on the direction of the wavenumber $\vec k$, and not only on its modulus $k\equiv \|\vec k\|$. Expanding ${\cal P}_{\mathcal{R}}(\vec k)$ in spherical harmonics, one obtains
\be \label{anisP}
{\cal P}_{\mathcal{R}}(\vec k)=\sum_{{\rm even} \, L}\sum_{M=-L}^L\, P_{\mathcal{R}}^{LM}(k)\, Y_{LM}(\hat k)\, ,
\ee
where $\hat k = \vec k / \| \vec k \|$. The sum over $L$ must be restricted to even values%
\footnote{Consequently, this parameterization cannot describe a dipolar $(L=1)$ modulation, like the one discussed in \cite{Planck:2018jri}. A different mechanism is required to obtain a sizable dipolar asymmetry in the CMB, for example large primordial non-Gaussianities \cite{Schmidt:2012ky, Agullo:2020fbw, Agullo:2020cvg, Agullo:2021oqk}.}
as a consequence of the homogeneity of the Bianchi~I geometry, which implies%
\footnote{The power spectrum is defined from the two-point function in Fourier space by the equation $\langle \mathcal{R}_{\vec k} \mathcal{R}_{\vec k'} \rangle = (2\pi)^3 \delta(\vec k+\vec k') \frac{2\pi^2}{k^3} \, {\cal P}_{\mathcal{R}} (\vec k)$. Recall that the presence of the delta function is due to the underlying homogeneity. This delta function implies that $\langle \mathcal{R}_{\vec k'} \mathcal{R}_{\vec k} \rangle = (2\pi)^3 \delta(\vec k'+\vec k) \frac{2\pi^2}{k^3} \, P_{\mathcal{R}}(-\vec k)$. But since $[\mathcal{R}_{\vec k}, \mathcal{R}_{\vec k'}]=0$ the two spectra must agree, ${\cal P}_{\mathcal{R}}(\vec k) = {\cal P}_{\mathcal{R}}(-\vec k)$.}
${\cal P}_{\mathcal{R}}(\vec k) = {\cal P}_{\mathcal{R}}(-\vec k)$. Therefore, the lowest anisotropic multipole is $L=2$, and it has the angular dependence of a quadrupole. We will show in Sec.~\ref{pred} that this quadrupole is the dominant anisotropic feature arising in bouncing alternatives to inflation, and consequently we focus on it from now on.

It is convenient to factorize the isotropic (monopolar) contribution, rewriting Eq.~\eqref{anisP} as
\be \label{anisPg}
{\cal P}_{\mathcal{R}}(\vec k) = \frac{1}{\sqrt{4\pi}} {\cal P}^{00}_{\cal R}(k) \left( 1 + \sum_{M=-2}^{M=2} g_{2M}(k) \, Y_{2 M}(\hat k) + \cdots \right) \, ,
\ee
where the factor $1/\sqrt{4\pi}$ originates from $Y_{00}(\hat k)=1/\sqrt{4\pi}$. The quantities $g_{2M}(k)$ describe the amplitudes of the quadrupolar contributions relative to the isotropic monopole, namely, ${\cal P}_{\cal R}^{2M}(k) = \frac{1}{\sqrt{4\pi}}\, g_{2M}(k) {\cal P}^{00}_{\cal R}(k)$. Note that the quadrupolar modulation can be scale-dependent if $g_{2M}(k)$ are allowed to depend on $k$.

The Planck collaboration searched for such a quadrupolar modulation in the CMB \cite{Planck:2018jri}. The possible scale dependence of $g_{2M}(k)$ was restricted to a power law of the form $g_{2M}(k)= \bar g_{2M}\, \left(\frac{k}{k_\star}\right)^q$, with possible values $q=0,\pm 1,\pm 2$, and where $k_\star=0.05\, {\rm Mpc}^{-1}$ is a reference scale; the case $q=0$ corresponds to a scale-invariant quadrupole. The Planck collaboration reports the best fit for the average amplitude $g_2\equiv \sqrt{\sum_M |\bar g_{2M}|^2/5}$ for these values of $q$ in Table~17 of Ref.~\cite{Planck:2018jri}. Of particular interest for this paper is the scale-invariant case $q=0$, for which the Planck collaboration reports
\be \label{g2-planck}
g_2^{\rm obs} = 7.62 \times 10^{-3} \, .
\ee
The uncertainty in this value is quantified in terms of the so-called $p$-value, defined as the percentage of computer-generated random simulations using an isotropic probability distribution with a value of $g_2$ at least as large as the observed one---roughly speaking, the chance that the quadrupole \eqref{g2-planck} could be observed in an isotropic universe. For a scale-invariant quadrupole with $q=0$, this $p$-value is reported to be $51.7$. In other words, about half of the realizations of an isotropic probability distribution for the CMB contains a value of the quadrupolar modulation as large as the observed one. Consequently, the statistical significance of $g_2^{\rm obs}=7.62\times 10^{-3}$ is much too small to claim a detection of primordial anisotropies. Nonetheless, this value is still useful since it provides an upper bound for how large the primordial quadrupolar modulation could be in our universe: any cosmological scenario that predicts a value for $g_2$ greater than \eqref{g2-planck} is ruled out. We will use this to constrain anisotropic shears in bouncing alternatives to inflation.

\section{Impact of anisotropies}
\label{dyn}

In this section we study quantitatively how anisotropies evolve in bouncing models of the early universe, both for the background geometry and for perturbations, and use the results to compute the primordial power spectrum. Our primary aim here is to understand the role of anisotropies during the contracting phase of the cosmos, just before the bounce, for different types of matter. We do not consider constraints from observations here; these will be discussed in the following section. 

For concreteness, we will assume a Bianchi~I background with the line element
\be
\dd s^2 = - \dd t^2 + a^2_1(t)\, \dd x_1^2 + a^2_2(t) \, \dd x_2^2 + a^2_3(t)\, \dd x_3^2,
\ee
where $a_i(t)$ are the three directional scale factors. In addition, we will focus here on three representative types of matter fields: radiation, stiff matter, and an ekpyrotic fluid. Anisotropies are measured by the shears $\sigma_i = H_i - H$, which compare the directional Hubble rates $H_i = \dot a_i / a_i$ to the Hubble rate $H = \dot a / a$ of the mean scale factor $a = (a_1 a_2 a_3)^{1/3}$; dots denote derivatives with respect to $t$. The three matter fields we consider have an equation of state $p = w \, \rho$ with a constant $w$, for radiation $w_r=\frac{1}{3}$, for stiff matter $w_s = 1$, and we take $w_e = 3$ for the ekpyrotic fluid. The way the energy density $\rho$ and the anisotropic shears $\sigma_i$ evolve under the influence of these fluids is well known%
\footnote{This scaling holds in general relativity. In our simulations, the bounce is caused by adding corrections to general relativity, and these corrections introduce deviations from the well-known scaling in general relativity; however, these deviations are only sizeable during a short interval around the bounce and do not significantly affect the scaling given here away from the bounce. The deviations from general relativity do not have any significant effect on our calculations except creating a bounce.}:
$\rho\propto a^{-3\,(1+w)}$, while $\sigma_i\propto a^{-3}$ and $\sigma^2 \equiv \sigma_1^2 + \sigma_2^2 + \sigma_3^2 \propto a^{-6}$; see App.~\ref{Bianchi} for details.

\subsection{General cosmological evolution}

We start by describing the different cosmological eras we include in the calculations; quantitative results derived from it are reported in the next subsections.

\bigskip

\noindent
{\bf Initial conditions}

We set initial conditions in a contracting universe well before the bounce, assuming the space-time metric can be approximated by a Bianchi~I background with linear perturbations.

For the background, we assume that the anisotropies are initially very small compared to the energy density. (In principle, initial conditions with large anisotropies are also possible but, as we shall see, this possibility is ruled out by CMB constraints; therefore, in the following we only consider initial conditions where the anisotropic shears are subleading, as required by observations.)

For the perturbations, we assume that the scalar modes (defined below) are nearly scale-invariant with an appropriate amplitude, while tensor modes are negligibly small and taken to be zero. By choosing these initial conditions, we are assuming that nearly scale-invariant scalar perturbations were generated at an earlier time in the contracting universe, for example during a phase of ekpyrosis or matter-dominated contraction. We also assume that, at the time we start the evolution deep in the contracting branch, perturbations are in an isotropic quantum state. This assumption is justified by the strong upper bounds we will obtain later in this paper for anisotropic shears in the Bianchi I metric; such constraints imply that, before the time we start the evolution, the universe was almost isotropic. This in turn implies that any potential anisotropic features the perturbations may have acquired in the previous evolution must be very small and can safely be neglected. In terms of the power spectrum given in \eqref{anisPg}, these initial conditions correspond to ${\cal P}^{00}_{\cal R}(k) \sim k^{-\epsilon}$ (with $0 < \epsilon \ll 1$) and ${\cal P}^{LM}_{\cal R}(k) = 0$ for any $L \neq 0$, or equivalently $g_{LM} = 0$ for all $L \neq 0$.

In some cases like matter-dominated contraction, nearly scale-invariant tensor modes will also be generated---this can lead to additional observational constraints on the model \cite{Quintin:2015rta}. Since our focus here is on the consequences of background anisotropies, for simplicity we set the initial spectrum for the tensor modes to vanish; importantly, in this way the analysis is applicable to ekpyrotic scenarios as well, which do not generate significant tensor modes. Note that in anisotropic space-times, the tensor and scalar modes are coupled, so assuming an initially vanishing tensor power spectrum is a conservative assumption, since anisotropies will tend to generate tensor modes; however, allowing for this effect will only strengthen observational constraints on anisotropies, and for the sake of simplicity we neglect any generation of tensor modes before the beginning of our numerical simulations. This assumption is further justified by the strong upper bounds we will obtain later in this paper for anisotropic shears; such constraints imply that, before the time we start the evolution, the interaction between tensor and scalar perturbations must have been extremely weak and can safely be neglected.

This summarizes the initial conditions for all of the runs we consider. We give the precise numbers for the initial conditions below, in the relevant sections.

\smallskip

\noindent
{\bf Pre-bounce evolution}

As already mentioned, for the matter content of the universe in the last stages of contraction before the bounce, we will explore three different possibilities: radiation, stiff matter, and ekpyrosis. These are perfect fluids with vanishing anisotropic stress and a constant equation of state $w = p/\rho$, with $w_r=\frac{1}{3}$ for radiation, $w_s=1$ for stiff matter, and we choose $w_e=3$ for ekpyrosis.

We assume that during the contracting phase general relativity holds with great accuracy, with deviations only becoming important in the immediate vicinity of the bounce. As a result, the usual Einstein equations can be used for both the background and perturbative dynamics during contraction.

For simplicity, we mimic the perfect fluid behaviour with a constant equation of state $w$ by using a scalar field $\phi$ with an appropriate potential \cite{Heard:2002dr, Mielczarek:2008qw}
\be \label{pot-cl}
V(\phi) = V_o \exp \left( - \sqrt{24 \pi (1+w)} \, \phi \right).
\ee
In Bianchi~I space-times, it is not strictly true that a scalar field with such a potential produces exactly the required equation of state, but due to the strong constraints on anisotropies from CMB observations, all of the Bianchi~I geometries that we consider have very small anisotropies, even at the bounce, so the scalar field can mimic a perfect fluid with constant $w$ to a very good accuracy: the deviation of the value of $p/\rho$ from a constant value equal to 1/3, 1 or 3 is smaller than 0.01\% in the runs we study.

Another reason to take a scalar field as matter is that there already exists a numerical code (see \cite{Agullo:2020wur}) that can numerically solve the equations of motion, given in \cite{Pereira:2007yy, Agullo:2020uii}, for cosmological perturbations on a Bianchi~I background minimally coupled to a scalar field---we will use this already existing code for our simulations. We leave for future work a similar calculation using instead the equations of motion for cosmological perturbations with perfect fluid matter on anisotropic backgrounds that are given in App.~A of \cite{Pitrou:2015iya}. Although this remains to be confirmed in future work, we expect that the main results will be qualitatively very similar whether using a scalar field or a perfect fluid as matter source, just as they are in isotropic cosmologies.

\smallskip

\noindent
{\bf The bounce}

To carry out our simulations it is necessary to introduce a mechanism to produce a cosmic bounce. This can be done either by adding an extra suitable matter field that violates the energy conditions of general relativity, or by introducing corrections to the gravitational field equations. Our goal is to learn how anisotropies grow in the contracting, pre-bounce phase, and consequently we are interested in a mechanism to create a bounce which affects as little as possible our conclusions. This can be achieved if the bounce provides a quick transition between contraction and expansion, fast enough that it does not significantly change the perturbations or modify the dynamics at times far before the bounce.

A convenient way to achieve a bounce in the Bianchi~I background is by using the modifications to the Einstein equations that appear in loop quantum cosmology (LQC) \cite{Ashtekar:2011ni, Agullo:2016tjh}, and which we summarize in App.~\ref{LQC}. LQC is known to be compatible with ekpyrosis \cite{Cailleteau:2009fv, Wilson-Ewing:2013bla} as well as the matter bounce \cite{Wilson-Ewing:2012lmx, Li:2020pww}, and the predictions of the models are to a large degree insensitive to the physics of the bounce \cite{Cai:2014zga}. LQC is also a convenient way to introduce a non-singular cosmic bounce for the following reasons. First, the modifications to general relativity depend on a tunable parameter $\rho_b$, which controls the value of the space-time curvature at the bounce (and, for small anisotropies, this can be translated into a bound on the energy density). In the context of quantum gravity, it is natural to expect $\rho_b$ to be of the order of the Planck scale, but in this article we will treat it as a free parameter. Second, to maintain a constant equation of state through the bounce, it is possible to appropriately modify the potential \eqref{pot-cl} to \cite{Mielczarek:2008qw}
\be
V(\phi) = \f{V_o \exp \left( - \sqrt{24 \pi (1+w)} \, \phi \right)}{1 + \f{V_o}{2(1-w)\rho_b} \exp \left( - \sqrt{24 \pi (1+w)} \, \phi \right)^2}~.
\ee
As expected, the standard potential is recovered in the limit $\rho_b \to \infty$. Third, the deviations from the Einstein equations are only relevant for a short period of time, of the order of the time scale determined by $\rho_b$ (in natural units). This guarantees that the conclusions we reach for cosmological perturbations do not strongly depend on the details of LQC, precisely because these details are important only for a very short time.

In summary, we use the equations of LQC as a phenomenological tool to produce a bounce in the Bianchi~I background space-time with a free physical parameter $\rho_b$ that can be used to select the energy scale of the bounce, but our analysis and conclusions are independent of the physical assumptions on which LQC rests. Essentially identical constraints would be obtained for other ways of generating a bounce, so long as the energy scale of the bounce is the same, and the effect of the new physics causing the bounce are wholly concentrated in a short time interval around the bounce.

In our calculations below we use $\rho_b = 4.54 \times 10^{-12}$ in Planck units. This number is far from the Planck scale and is singled out by demanding that the amplitude of the scalar power spectrum agrees with CMB observations for the radiation-dominated runs we study. For both stiff and ekpyrotic matter, and in general for fluids with $w \ge 1$, the amplitude of the scalar power spectrum is almost insensitive to the choice of $\rho_b$, since perturbations freeze on super-Hubble scales, and we could choose other values of $\rho_b$. However, we choose to use the same value, $\rho_b = 4.54 \times 10^{-12}$, in order to facilitate the comparison between the three matter contents we consider.

Finally, for the perturbations we continue to use the equations of motion of general relativity during the bounce. We make this choice for simplicity, and also because it is the most conservative choice we can make. Although the new physics causing the bounce could in principle introduce new effects at the perturbative level as well and modify the equations of motion for the perturbations, any such effects will only be important for the short time it takes for the bounce to occur and it seems unlikely that they will significantly modify the main results.

In the numerical simulations described below in more detail, we find that the spectrum of the perturbations does not change significantly during the bounce, as expected given the discussion here.

\smallskip

\noindent
{\bf The expanding branch}

In FLRW isotropic spacetimes, curvature perturbations with super-Hubble wavelengths remain constant in an expanding universe. In this phase, perturbations are only sensitive to the redshift effect, which measures how much the universe expands in total, but they are independent of the Hubble rate and other details of the homogeneous and isotropic matter content of the universe. Although this is no longer true for Bianchi~I geometries in general, we are only studying space-times where the anisotropies are very small and are bounded by strong constraints in the CMB. In this particular case, deviations from FLRW are small. For the evolution of super-Hubble perturbations in the expanding phase they are negligible and these super-horizon modes remain constant (to an excellent approximation) in the expanding branch. For this reason, our conclusions are insensitive to the matter content of the universe after the bounce, as long as there are no extra sources of anisotropies (for example, fluids with large anisotropic stresses). In view of this, and also for simplicity, we assume the universe is dominated by radiation right after the bounce. We make this choice because the early universe must eventually become radiation-dominated, and the simplest choice is to transition to radiation right after the bounce. We ensure that the transition at the bounce to radiation domination is such that both the background metric and perturbations are continuous and differentiable.

\subsection{Anisotropic background geometry}

Next, we study the evolution of the anisotropies in contracting Bianchi~I space-times, for the three representative matter contents already mentioned, namely radiation $w_r=1/3$, a stiff fluid $w_s=1$, and an ekpyrotic fluid with $w_e=3$, denoting these three matter fields by the indices $r$, $s$ and $e$ respectively.

Specifically, we perform numerical simulations and monitor the evolution of the directional scale factors $a_i(t)$, from which we can compute anisotropic shears and the energy density. We present three representative simulations, one for each matter content. In order to compare the simulations for these different choices of matter in a meaningful manner, we start the evolution at some arbitrary time in the contracting phase of the universe and use the same initial conditions for the metric and the energy density of the matter content for the three simulations; in this way the differences in the subsequent evolution can only be attributed to the matter content and not to the initial conditions.

We fix the initial conditions in the following way. As described in App.~\ref{Bianchi}, a complete set of initial data is given by specifying the value of five parameters, namely (i) the energy density $\rho$, (ii) the equation of state $w$, (iii) the total shear squared $\sigma^2$, (iv) the sign of the mean Hubble rate and (v) the parameter $\Psi$ that controls how the anisotropies are distributed among the three principal directions. The values we use for the numerical simulations presented in this section are given in Table~\ref{indata}. Note that we assume that the anisotropies are initially very small compared to the energy density, $\sigma_{in}^2/16\pi\ll \rho_{in}$. As discussed above, we set $\rho_b = 4.54 \times 10^{-12}$ (in Planck units) for the energy density at the bounce in all our simulations.

\begin{table}
\centering
\begin{tabular}{|c|c|c|c|}
\hline
$\rho_{in}$ & $\sigma_{in}^2$ & $\rm{sign}(H_{in})$ & $\Psi$ \\ \hline 
$4.96\times 10^{-18}$ & $4.15 \times 10^{-26}$ & ${\rm negative}$ & $0$ \\
\hline
\end{tabular}
\caption{Initial data for the background geometry. The energy density $\rho_{in}$ and shear $\sigma_{in}^2$ are expressed in Planck units, and we use three different choices for $w$, $w_r=1/3$, $w_s=1$ and $w_e=3$. We have explored different values of $\Psi$ in our simulations and the main results are insensitive to the choice of $\Psi$. For concreteness we set $\Psi=0$ for the simulations we present here.}
\label{indata}
\end{table}

\begin{figure}[t]
{\centering
\includegraphics[width = 0.48\textwidth]{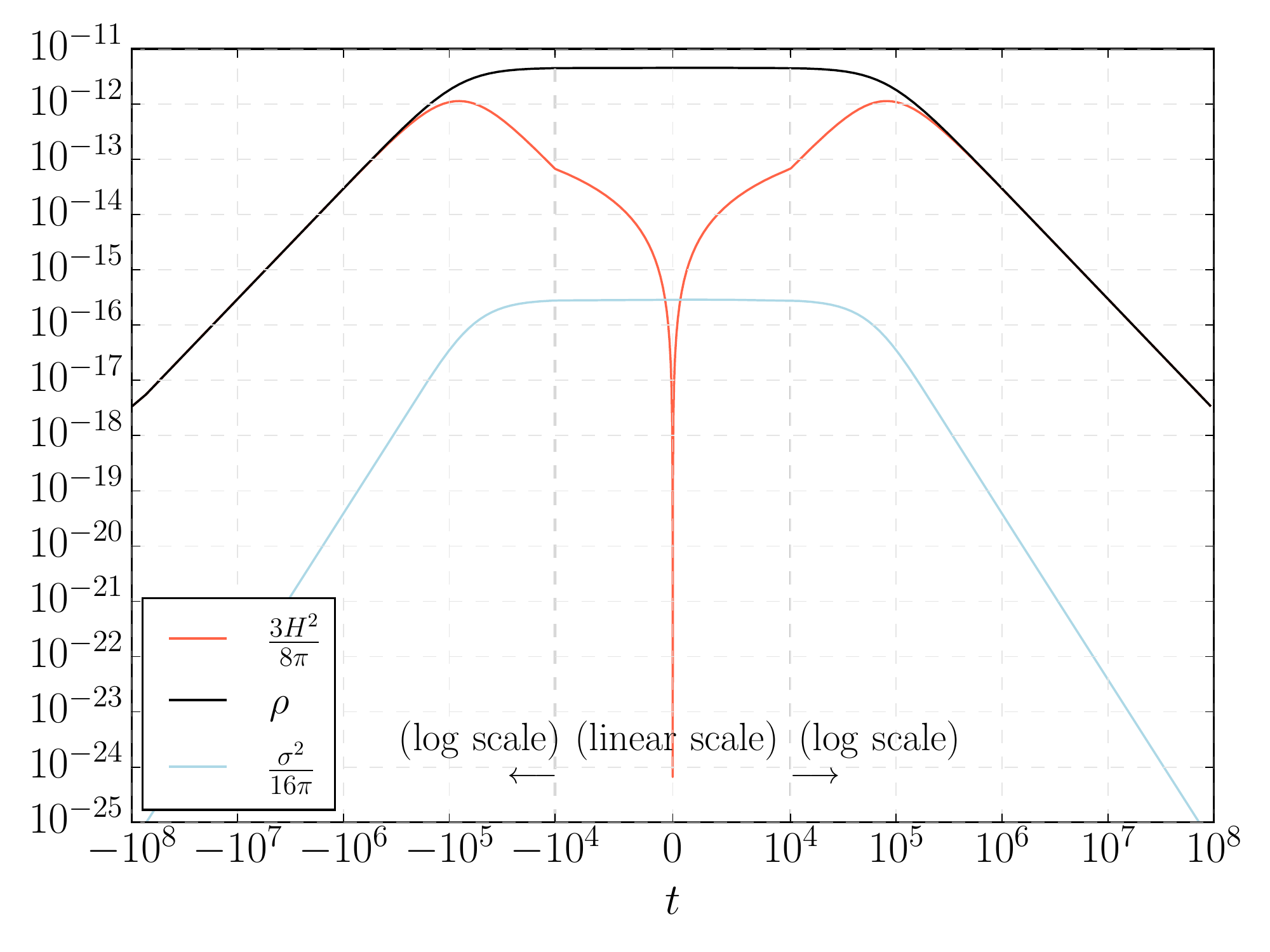}
\includegraphics[width = 0.48\textwidth]{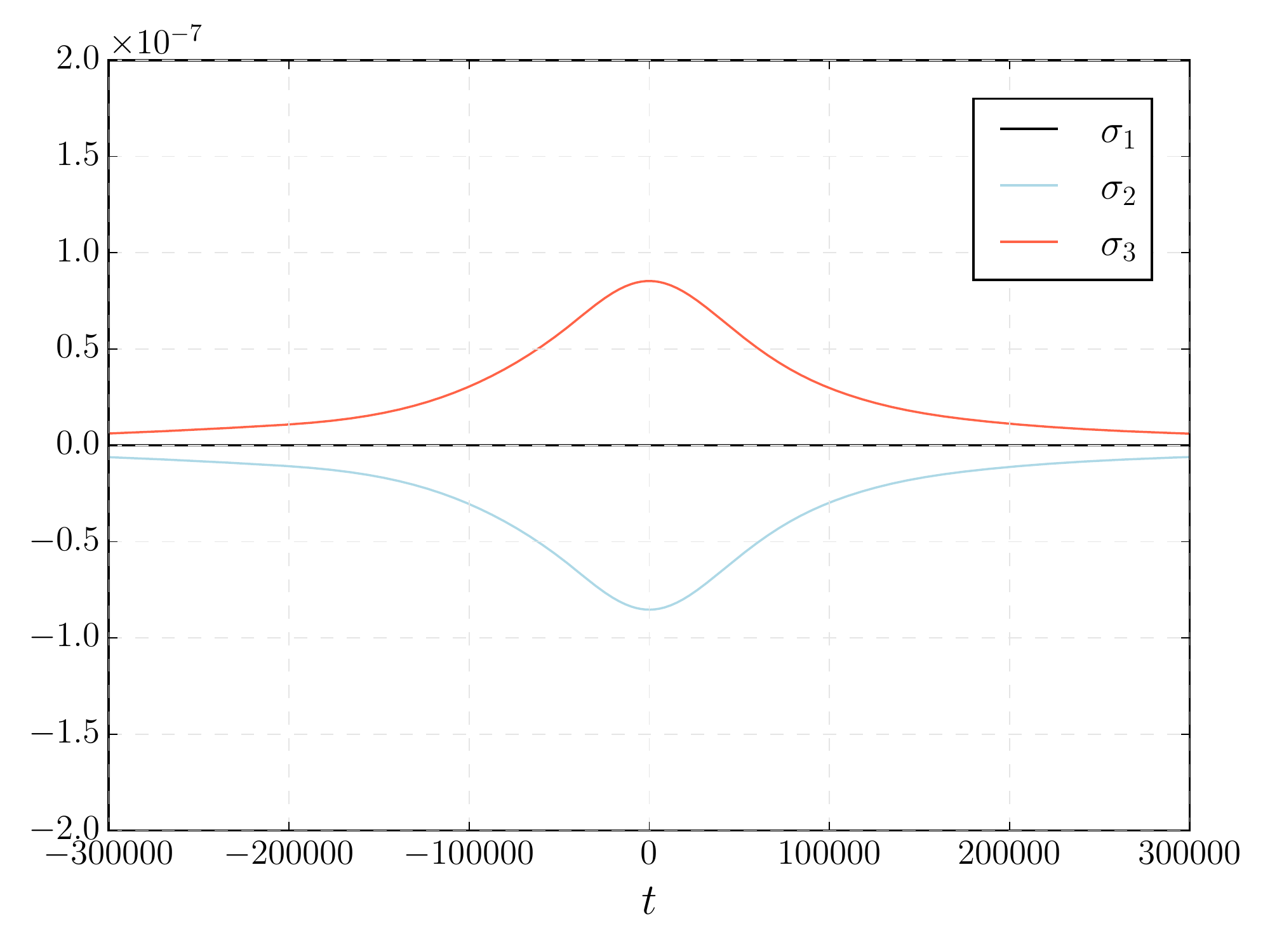}
}
\caption{These plots show results obtained for a radiation-dominated bouncing universe with the initial data specified in Table~\ref{indata}. The results for the other matter fields are qualitatively similar. \\
\textit{Left panel:} Evolution of $\rho(t)$ and $\sigma^2(t)/16\pi$ in cosmic time $t$. They both reach a maximum value at the bounce at $t=0$, and in this simulation $\sigma_{in}^2/16\pi \ll \rho_{in}$ at all times. We also plot $3H^2(t)/8\pi$, which is equal to $\rho$ away from the bounce where general relativity holds, but deviates from it at the bounce. Note that the scale of the vertical axis is always logarithmic, while the scale for the horizontal axis is logarithmic for $|t| > 10^4$ but linear for $|t| < 10^4$. \\
\textit{Right panel:} Evolution of the three anisotropic stresses $\sigma_i(t) = H_i(t) - H(t)$, all three anisotropic stresses reach their maximum (in magnitude) near the bounce point. The scales of both axes are linear. Note that $\sigma_1$ is always zero; this is due to the choice of $\Psi=0$.} \label{fig1}
\end{figure}

\begin{figure}[t]
{\centering
\includegraphics[width = 0.49\textwidth]{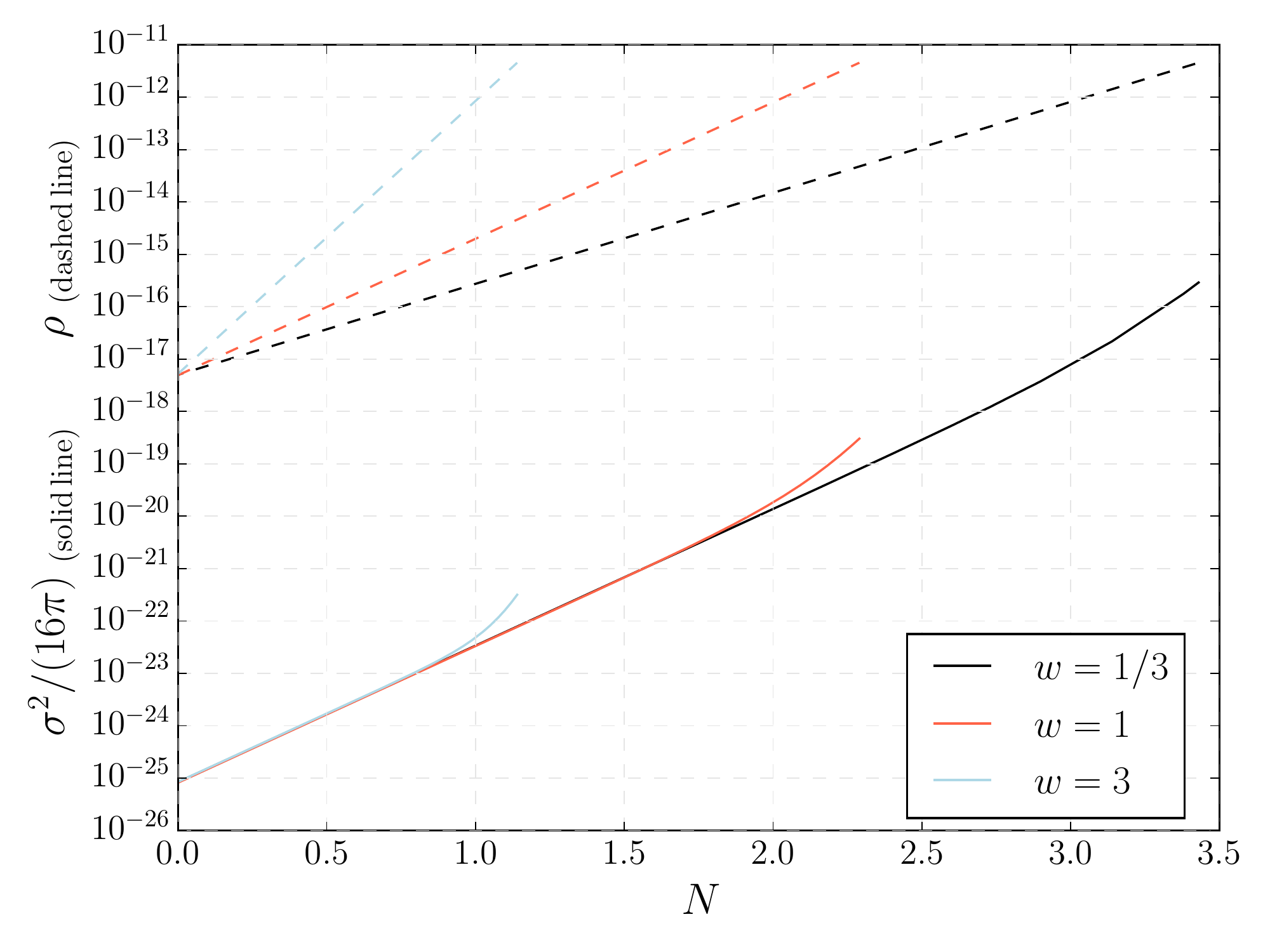}
\includegraphics[width = 0.49\textwidth]{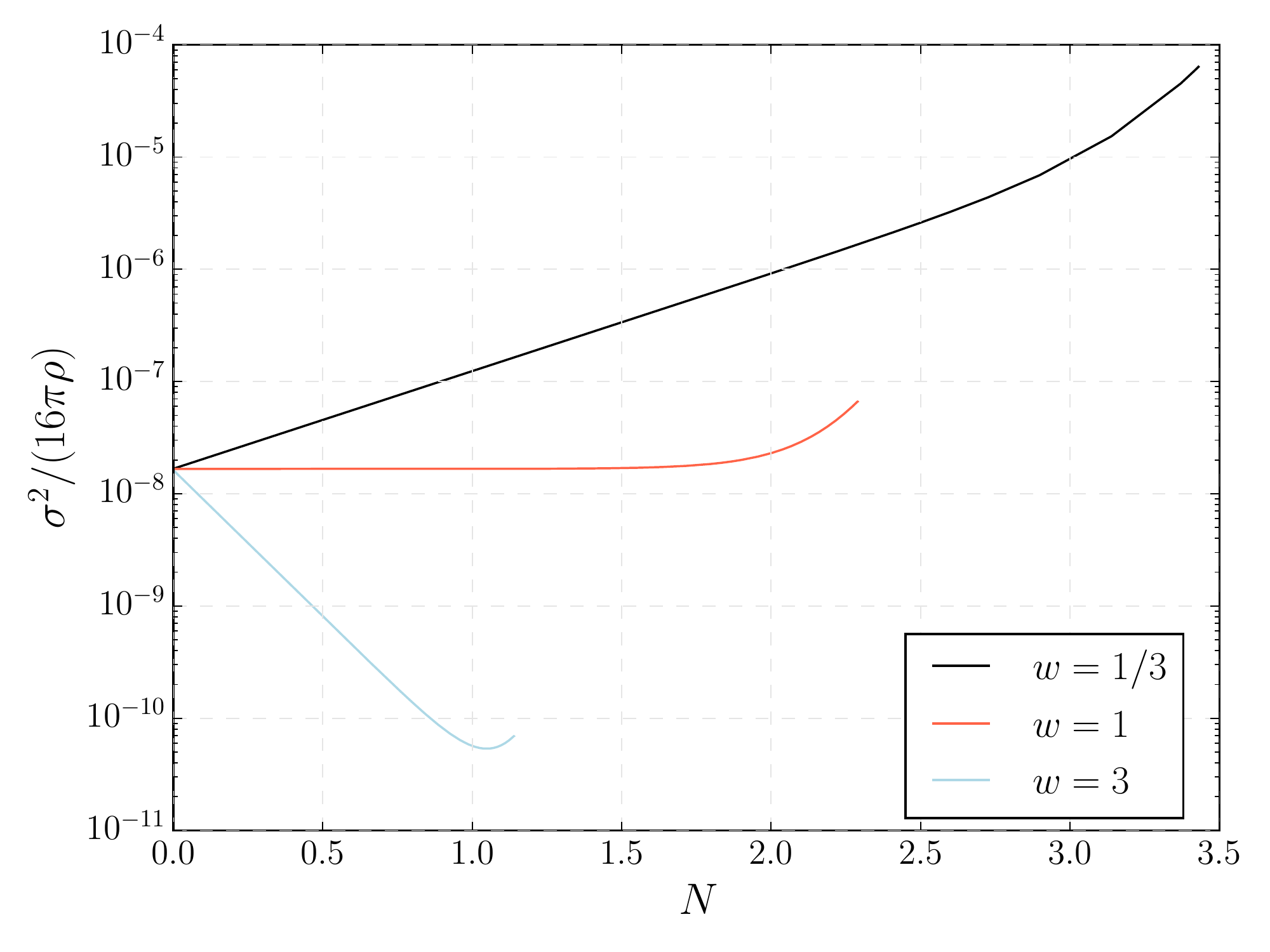}}
\caption{These plots show the growth in the energy density of the matter content and in the anisotropies as a function of the number of $e$-folds of contraction $N$ from the initial time to the bounce for different equations of state. For the three matter contents under consideration, we choose the same value of the energy density initially, as given in Table~\ref{indata}, and we set $\rho_b= 4.54 \times 10^{-12}$ in Planck units. The total number of $e$-folds from the initial conditions to the bounce is $N_e=1.14$ for ekpyrosis (with $w=3$), $N_s=2.29$ for stiff matter, and $N_r=3.43$ for radiation. \\
\textit{Left panel:} This plot shows the energy density of the matter field, which grows as $a^{-3(1+w)}$, denoted by dashed lines. Since the bounce happens at a fixed curvature scale (directly related to $\rho$ since anisotropies are small in these simulations), if the equation of state $w$ is smaller, then more $e$-folds of contraction are required to reach the bounce. The plot also shows the anisotropies in solid lines; these grow as $a^{-6}$ (except for a short time close to the bounce where they grow more rapidly due to departures from general relatively), independently of the matter field. Anisotropies grow at the same rate in all three cases, but will grow for a longer time if there are more $e$-folds of contraction. Therefore, the anisotropies at the bounce will be larger for a smaller equation of state. \\
\textit{Right panel:} This plot shows the ratio $\sigma^2/(16\pi\rho)$. In a contracting universe, this ratio grows for $w<1$, is constant for $w=1$, and decreases for $w>1$ (except potentially in the vicinity of the bounce where there are departures from general relativity). For radiation and stiff matter, this ratio reaches a maximum value at the bounce, respectively ${}^{(s)}\sigma^2_b / (16\pi\rho_b) = 6.66 \times 10^{-8}$ and ${}^{(r)}\sigma^2_b / (16\pi\rho_b) = 6.37 \times 10^{-5}$, while for the ekpyrotic case this ratio reaches a minimum near the bounce, ${}^{(e)}\sigma^2_{min} / (16\pi\rho_{min}) = 5.34 \times 10^{-11}$ at $N=1.04$, and then (due to departures from general relativity around the bounce) slightly grows until before reaching the bounce at $N_e=1.14$.
}
\label{fig2}
\end{figure}

With this initial data, we numerically solve the equations for the background metric and the scalar field using the equations of motion derived from the Hamiltonian constraint~\eqref{lqc-ham}, as described in App.~\ref{LQC}. This is a set of coupled, second-order ordinary differential equations for the directional scale factors $a_i(t)$ and the scalar field mimicking the perfect fluid. We solve the dynamics using the explicit embedded Runge-Kutta Prince-Dormand $(8, 9)$ method of the GNU scientific library; the main results are summarized in Figs.~\ref{fig1} and \ref{fig2}.

Some results for the radiation-dominated case are shown in Fig.~\ref{fig1}. The left panel allows us to compare $\rho(t)$ and $\sigma^2(t)/16\pi$, while the right panel shows the three anisotropic stresses $\sigma_i(t) = H_i(t) - H(t)$. Given the initial conditions with a very small shear, the shear remains subdominant compared to $\rho$ at all times including the bounce, but it clearly grows more rapidly than $\rho$ in the left panel, and reaches a maximum at the bounce, as can be seen in both the left and right panels. While these results are for a bouncing Bianchi~I space-time dominated by radiation, the other matter fields we have considered give qualitatively very similar results, with the only difference being that the comparative rate of growth between the anisotropic shear and the matter energy density changes, as shall be explained next.

Then, we compare the evolution of the matter energy density for radiation $w_r=1/3$, a stiff fluid $w_s=1$, and an ekpyrotic fluid with $w_e=3$; we use the number of (contracting) $e$-folds, defined as $N = \ln(a_{in} / a_{fin})$ in terms of the mean scale factor, as the time variable. The evolution of the energy density $\rho(t)$ from the initial time until the time of the bounce is shown in dashed lines in the left panel of Fig.~\ref{fig2} for each choice of matter. Since the bounce happens at the energy scale $\rho_b$, it is not surprising that in each case $\rho$ increases following a power law with respect to $a(t)$ until $\rho = \rho_b$, at which point the bounce occurs. Of course, $\rho \sim a^{-3(1+w)}$ increases much more rapidly in a contracting space-time when the equation of state $w$ is increased, so given identical initial conditions a universe with an ekpyrotic field will bounce in fewer e-folds than in the case of stiff matter, which in turn bounces in fewer e-folds than a radiation-dominated universe.

The continuous and dashed lines in the left panel of Fig.~\ref{fig2} show the evolution of $\sigma(t)$ and $\rho(t)$, respectively, while the right panel shows the ratio $\sigma^2/(16\pi \, \rho)$ for each of the three matter fields we consider. These plots show precisely what one would expect from expressions $\rho \sim a^{-3\,(1+w)}$ and $\sigma^2 \sim a^{-6}$, combined with the information that the bounce occurs when $\rho = \rho_b$. First, since the anisotropies grow at the rate $\sigma^2 \sim a^{-6}$ independent of the matter field (in the absence of viscosity which can modify this relation \cite{Misner:1967zz, Ganguly:2019llh, Ganguly:2021pke}), the curves for $\sigma^2$ are (before deviations from general relativity arise) identical, the only difference is the number of $e$-folds before the bounce occurs. Since the anisotropies stop growing at the bounce (and decrease during the expanding phase), they will reach a greater maximum for radiation than for ekpyrosis, as can be seen in the left panel of Fig.~\ref{fig2} (the curves end at the point where the bounce occurs).

A simple although important point is that anisotropies, as captured by $\sigma^2$, can grow significantly in a contracting universe for any type of matter content---the relevant quantity is the number of $e$-folds of contraction, not the equation of state $w$ of the matter field.

Second, the results shown by the dashed and solid curves in the left panel of Fig.~\ref{fig2} are combined in the right panel of Fig.~\ref{fig2}, where the ratio $\sigma^2/(16\pi \, \rho)$ is plotted, showing that this ratio grows for a radiation-dominated contracting universe, remains nearly constant for stiff matter, and decreases for ekpyrotic contraction. These are well-known results, commonly used to argue that an ekpyrotic fluid dilutes anisotropies: if initially the ratio $\sigma^2/(16\pi\, \rho)$ is small, it will become even smaller during ekpyrotic contraction. We will argue below, using perturbations, that this condition alone is not sufficient to meet all observational constraints from the CMB.

Note that near the bounce, the scaling behavior expected from general relativity for the shear $\sigma^2$ is modified because the departures from Einstein's theory become large (as they must to cause a bounce). For instance, we see that around the bounce $\sigma^2$ grows faster than $ a^{-6}$, while $\rho$ keeps its classical behavior. This implies, for instance, that the ratio $\sigma^2/(16\pi\rho)$ is not constant around the bounce for stiff matter. Even for ekpyrotic matter, the ratio $\sigma^2 / (16 \pi \rho)$ can actually grow for a short time close to the bounce if $w<5$. Nonetheless, this deviation from general relativity happens during a very short time interval, and is therefore a small effect that does not have a significant impact on our main conclusions.

\subsection{Impact of anisotropies on perturbations}
\label{subsec:perts}

The dynamics of cosmological perturbations on a Bianchi~I homogeneous background is considerably more complicated than on FLRW backgrounds. We will follow the analysis of Refs.~\cite{Agullo:2020uii, Agullo:2020iqv}, and summarize in App.~\ref{Bianchi} the information needed for our calculations.

When the matter source is a scalar field with no anisotropic stresses, cosmological perturbations in Bianchi~I geometries are described by three fields $\Gamma_{\mu}$, with $\mu=0,1,2$; these are the generalization of the familiar scalar and tensor perturbations on FLRW backgrounds to Bianchi I geometries. When the anisotropic shears become small at late times in the expanding branch and the universe isotropizes, $\Gamma_{0}$ becomes proportional to the co-moving curvature perturbation, $\Gamma_{0}=\sqrt{32\pi }\, (z/a) \cdot \mathcal{R}(\vec{k})$, where $z = a \, \dot \phi / H$, while $\Gamma_1$ and $\Gamma_2$ reduce to the $+,\times$ polarizations of tensor modes, respectively. The equations of motion for the Fourier components of the $\Gamma_{\mu}$ fields are \cite{Agullo:2020uii}
\be \label{eqginper}
\ddot \Gamma_{\mu}(\vec k)+3\, H\, \dot \Gamma_{\mu}(\vec k)+\frac{k^2}{a^2}\, \Gamma_{\mu}(\vec k)+\frac{1}{a^2}\, \sum_{\mu'=0}^2\, {\cal U}_{\mu\mu'} (\hat k, t) \, \Gamma_{\mu'}(\vec k)=0\, ,
\ee
where dots denote derivatives with respect to cosmic time $t$, and
\be
k^2(t)\equiv a^2(t)\, \left(\frac{k_1^2}{a_1^2(t)}+\frac{k_2^2}{a_2^2(t)}+\frac{k_3^2}{a_3^2(t)}\right).
\ee
Note that $k(t)$ now depends on time, unlike on a FLRW background; this is because modes with different $\hat k$ will be red-shifted at rates depending on their directional scale factor $\hat k \cdot \vec a$, where $\vec a$ is the vector with components $(a_1, a_2, a_3)$. In an anisotropic universe, the modes with equal modulus $k$ today did not necessarily have the same modulus at earlier times---this is an effect that must be tracked when solving the equations of motion.

In Eqs.~\eqref{eqginper}, the functions ${\cal U}_{\mu\mu'}(\hat k, t)$ play the role of effective potentials, and their expressions are given in App.~\ref{app.perts}. Importantly, the potentials ${\cal U}_{\mu\mu'}(\vec k, t)$ depend on the anisotropic shears $\sigma_i$ as well as on the unit vector $\hat k$; this is in addition to their dependence on isotropic quantities like $H$ and contributions from matter fields like $\dot\phi$. Note however that each ${\cal U}_{\mu\mu'}(\hat k, t)$ does not depend on the modulus $k$: each potential is the same for all wave vectors pointing in a given direction.

When the anisotropic shears $\sigma_i$ are non-zero, the equations of motion for $\Gamma_\mu$ have two important properties that distinguish them from the isotropic case: (i) the three fields $\Gamma_{\mu}(\vec k)$ are coupled because the potentials ${\cal U}_{\mu\mu'}(\hat k, t)$ do not vanish for $\mu \neq \mu'$, and (ii) the potentials ${\cal U}_{\mu\mu'}(\hat k, t)$ depend on $\hat k$ (but are independent of $k$), through $\hat k \cdot \vec \sigma$ where $\vec \sigma$ is the vector with the components $(\sigma_1, \sigma_2, \sigma_3)$. The existence of these two basic properties is easily understood (although the explicit form of the ${\cal U}_{\mu\mu'}$ potentials is complicated) and these properties can have important observational consequences.

The first property arises because rotations are no longer a symmetry in anisotropic space-times. In isotropic FLRW space-times, the scalar and the two tensor modes decouple precisely due to the rotational symmetry; see for example App.~A2 of \cite{Baumann:2009ds} for a proof of this. For an anisotropic background, although it remains possible to split the three perturbative degrees of freedom into one scalar and two tensor modes, in the absence of a rotational symmetry these three degrees of freedom are coupled. A consequence of this coupling is that it will typically introduce cross-correlations between scalar and tensor perturbations.

The second property is due to the cosmological perturbations depending on the directional Hubble rates, not only the mean Hubble rate. For example, two modes with wave-numbers of the same magnitude but pointing in different directions experience different potentials, and end up with different amplitudes. This directional dependence will generate anisotropies in the CMB, in particular a non-zero quadrupole moment $g_2$; this second property is the origin of the quadrupole moment which provides the strongest bounds on anisotropies from CMB data.

Note that anisotropies in perturbations are acquired continuously throughout the evolution---even if the shears always remain small, anisotropic features in the perturbations can nonetheless become significant as a result of the integration over time. In particular, even if $\sigma^2/(16 \pi \rho)$ decreases during the contraction, anisotropic features in perturbations can grow. Further, note that although the relative importance of anisotropies for the homogeneous background is given just by the ratio $\sigma^2/(16 \pi \rho)$, in contrast perturbations are sensitive to other components of the shear as well. As can be seen in App.~\ref{app.perts}, the potentials ${\cal U}_{\mu\mu'}(\vec k, t)$ depend on the components of the shear in a complicated manner, and some of these quantities can increase in time even when $\sigma^2/(16 \pi\rho)$ decreases. Because of this complexity, it is difficult to estimate how anisotropies in perturbations behave without numerical solutions on a case by case basis.

Finally, our main goal in this subsection is to compute the multipolar components of the primordial scalar power spectrum $g_{LM}(k)$, as defined in Eq.~\eqref{anisPg}, at late times after the bounce. Note that by definition each $g_{LM}(k)$ depends only on the modulus of $\vec k$; the dependence on the direction is fully captured in $Y_{L M}(\hat k)$. The primordial power spectrum $P_{\mathcal{R}}(\vec k)$ is anisotropic if any of the functions $g_{LM}(k)$ are different from zero for $L \geq 2$. The $k$ dependence of $g_{LM}(k)$ determines the amplitude of anisotropies at different scales in the CMB; for example, if $g_{LM}(k)$ is independent of $k$, then the anisotropies in the perturbations are scale-invariant.

The evolution of perturbations requires a choice of initial data. We will assume that, at the time at which we initiate our simulations, the wave-numbers of interest for the CMB are all super-Hubble (i.e., $k H / a \ll 1$), scalar perturbations have an almost scale-invariant power spectrum, and tensor perturbations are vanishingly small.

As an aside, note that in this paper we are not concerned with how the primordial perturbations are generated. Rather, we want to use perturbations as probes of anisotropies. This is why we simply assume that the primordial perturbations (nearly scale-invariant with a slight red tilt) have already been generated at an instant $t_{in}$ by means of one of the well-known mechanisms in bouncing alternatives to inflation, whether during a period of ekpyrosis \cite{Finelli:2002we, DiMarco:2002eb, Lehners:2007ac} or of matter contraction \cite{Wands:1998yp, Cai:2014jla}. We focus our investigations on the imprints that anisotropies induce, assuming that perturbations have already been generated by some appropriate mechanism.

More specifically, we assume that scalar perturbations, at the initial time $t_{in}$, have a power spectrum
\be \label{init-P}
{\cal P}_{\mathcal{R}}(t_{in},\vec k)=A_{in} \left(\frac{k}{k_{ref}}\right)^{n_s-1}\, ,
\ee
where $A_{in}$ is the value of the amplitude at a reference wavenumber $k_{ref}$, and $n_s$ is the usual spectral index with a small red tilt. For each scenario, $A_{in}$ is chosen to match the observed amplitude of the temperature-temperature correlations in the CMB, and in all cases $n_s-1=-0.036$.

With this initial condition the scalar perturbations are initially in an exactly isotropic state, while the background is nearly isotropic. Hence, all anisotropic features in the perturbations will be generated during the evolution, sourced by the anisotropic background at times later than $t_{in}$. Note that at $t_{in}$, since there are (small) anisotropies in the background it would be reasonable to expect (small) anisotropic features in the perturbations as well, rather than the perturbations being exactly isotropic as in Eq.~\eqref{init-P}. However, any anisotropic features in the perturbations should be small (given the almost-isotropic background space-time) and can safely be neglected---i.e., adding small anisotropic features to the initial conditions for the scalar perturbations will just give slightly stronger constraints on anisotropies.

For tensor modes, for simplicity we set $\Gamma_1(t_{in}) = \Gamma_2(t_{in}) = 0$, as would be expected for example in ekpyrosis models. In principle, any initial condition respecting the observational constraint from the CMB on the tensor-to-scalar ratio that $r < 0.036$ \cite{BICEPKeck:2021gln} is possible, although satisfying this bound may be a challenge for matter bounce scenarios \cite{Quintin:2015rta}. Setting the tensor modes to initially vanish will not significantly affect the predictions for scalar modes (and in particular for its quadrupole moment $g_2$) because although anisotropies couple the three $\Gamma_\mu$ fields, this coupling is driven by the anisotropies which remain small and so this effect is subleading (note also that due to this coupling $\Gamma_0$ will be a source for $\Gamma_1$ and $\Gamma_2$, which will not remain zero). Nonetheless, if one wishes to calculate anisotropic features in tensor perturbations (and cross correlations involving the tensor modes) in anisotropic bouncing cosmologies, the results will depend on the initial conditions for the tensor modes.

Given these initial conditions, we solve Eqs.~\eqref{eqginper} for the perturbations, using the values of $a_i(t)$, $a(t)$ and $H(t)$ derived in the previous subsections for each of the three types of matter fields sourcing the Bianchi~I geometry. We focus on the range of wave-vectors $\vec k$ that can be observed in the CMB, namely $k/k_* \in [0.002,4]$ (at the present time) with $k_*$ the wave-number whose physical value today is $0.05\, {\rm Mpc}^{-1}$. Contrary to the isotropic case, though, it is necessary to explore different directions of $\vec k$ and this requires significantly more computational power.

Concretely, for each simulation we express $\vec k$ using polar coordinates, in terms of its modulus $k$ and polar angles $\theta_{k}$ and $\phi_k$. We then discretize $k$ in steps of constant logarithmic size, and for each $k$, the polar angles take values on a $20\times40$ lattice with uniform spacing in the coordinates $u_k=\cos \theta_k$ and $\phi_k$, that covers the upper half of the unit sphere. The power spectrum in the lower half of the unit sphere is computed from the upper half by using the following properties of the power spectra: ${{\cal P}}_{\mu\mu'}(\vec{k})={{\cal P}}_{\mu'\mu}(-\vec{k})$, $\bar{{\cal P}}_{\mu\mu}(\vec{k})={{\cal P}}_{\mu\mu}(-\vec{k})$, $\bar{{\cal P}}_{01}(\vec{k})={{\cal P}}_{10}(-\vec{k})$, $\bar{{\cal P}}_{02}(\vec{k})=-{{\cal P}}_{20}(-\vec{k})$, and $\bar{{\cal P}}_{12}(\vec{k})=-{{\cal P}}_{21}(-\vec{k})$ (where the bar indicates complex conjugation); invariance under parity further implies ${{\cal P}}_{01}(\vec{k})={{\cal P}}_{01}(-\vec{k})$, ${{\cal P}}_{02}(\vec{k})=-{{\cal P}}_{02}(-\vec{k})$ and ${{\cal P}}_{12}(\vec{k})=-{{\cal P}}_{12}(-\vec{k})$; see \cite{Agullo:2020uii,Agullo:2020iqv} for a derivation of these properties%
\footnote{In short, they are derived from the fact that under Hermitian conjugation $\Gamma_0^{\dagger}(\vec k)= \Gamma_0(-\vec k)$ and $\Gamma_1^{\dagger}(\vec k)=\Gamma_1(-\vec k)$, while instead $\Gamma_2^{\dagger}(\vec k)=-\Gamma_2(-\vec k)$ (note the extra minus sign); and under parity $\Gamma_{\mu}(\vec k)\to \Gamma_{\mu}^{\dagger}(\vec k)$ for $\mu=0,1,2$ \cite{Agullo:2020uii}.}.
With these symmetries, it is sufficient to only calculate ${\cal P}_{00}(\vec k)$, ${\cal P}_{01}(\vec k)$, ${\cal P}_{02}(\vec k)$, ${\cal P}_{11}(\vec k)$, ${\cal P}_{12}(\vec k)$ and ${\cal P}_{22}(\vec k)$ for $\hat k$ in the upper half of the unit sphere; all remaining ${\cal P}_{\mu\nu}(\vec k)$ can be derived from these six using the properties listed above.

The calculation of each individual wave-vector $\vec k$ takes about half a minute, although the exact time depends on the concrete value of $k$ and the strength of the anisotropies. We run the evolution of all wave-numbers in parallel, and in a machine using 60 cores the entire calculation takes about three days. Once the power spectra ${{\cal P}}_{\mu\mu'}(\vec{k})$ are computed, we expand them in spin-weighted spherical harmonics to extract their angular multipole components \cite{Agullo:2020uii}.

\begin{table}
\centering
\begin{tabular}{|c|c|c|c|c|}
\hline
& Observational upper bound & Radiation & Stiff matter & Ekpyrosis ($w=3$) \\ \hline
$g_2$ & $7.62 \times 10^{-3}$ & $6.00 \times 10^{-3}$ & $2.65 \times 10^{-4}$ & $3.53 \times 10^{-5}$ \\
\hline 
\end{tabular}
\caption{Numerical results for the amplitude of the scale-invariant quadrupolar component of the primordial power spectrum using a background geometry corresponding to the initial data specified in Table \ref{indata}; and comparison with the upper bound obtained by the Planck Collaboration \cite{Planck:2018jri}.}
\label{g2} 
\end{table}

The leading order contribution to the $g_{LM}$ is a quadrupolar term with $L=2$; we have verified that higher-order terms $g_{LM}$ with $L \ge 4$ are at least an order of magnitude smaller for the radiation and stiff matter cases, and although we expect a similar result for the ekpyrotic case, numerical limitations do not allow us to accurately calculate higher-order $g_{LM}$ (a finer angular lattice in Fourier space is needed to resolve the $g_{LM}$ with $L \ge 4$ in this case). Due to this, we only report results for $g_2$ here.

The results of these numerical simulations show two important features worth emphasizing. First, as expected, we observe that larger anisotropies in the background generate larger anisotropies in the perturbations. Second, our numerical calculations reveal that the multipoles $g_{LM}(k)$ of the primordial power spectrum evaluated after the bounce are exactly scale invariant. Since all these coefficients are exactly zero for the initial state we use, the value of $g_{LM}(k)$ after the bounce must be attributed to the evolution. The scale invariance of the multipoles $g_{LM}(k)$ is an important result of our analysis and, although surprising at first, we provide below an explanation based on the form of the equations of motion. Given these results, the relevant observational constraint is the scale-invariant $q=0$ upper bound $g_2^{\rm obs} = 7.62 \times 10^{-3}$ \cite{Planck:2018jri}, as discussed in Sec.~\ref{constCMB}. For the initial conditions given above, the results of the numerical simulations for $g_2$ are summarized in Table~\ref{g2}. Notice that, even for the small initial value of the shear squared $\sigma^2_{in}$ considered in these calculations of $\sigma^2_{in}/(16\pi\, \rho_{in}) \sim 10^{-8}$, the value of $g_2$ is not far from the observational constraint.

As expected, for the same $\sigma^2_{in}$, the value of $g_2$ is larger for radiation, intermediate for a stiff fluid, and smaller for an ekpyrotic fluid, showing that an ekpyrotic fluid can indeed alleviate the problem with anisotropies in bouncing models. Nonetheless, the value of $g_2$ for ekpyrotic matter (with $w=3$) is only two orders of magnitude smaller than for radiation, even when the corresponding values of the shear squared at the bounce differ by six orders of magnitude (as shown in the left panel of Fig.~\ref{fig2}). Even for ekpyrotic cosmologies, observational constraints can be quite strong.

The predicted scale-invariance of $g_2$ is a key result. Its origin is due to the two facts that (i) the initial conditions are chosen, motivated by the assumption of an earlier (nearly) isotropic phase of ekpyrotic or dust contraction, such that $g_2 = 0$ initially, and (ii) that the observationally relevant modes $\vec k$ for the CMB are all super-Hubble ($k/a\ll H$) in the late contraction (and bounce) phase of the universe when anisotropies are non-negligible---in the limit $k/a\ll H$ the equations of motion for $\Gamma_\mu$ become independent of the modulus $k$, and the only remaining dependence is on the direction $\hat k$ which appears in the potentials ${\cal U}_{\mu\mu'}(\hat k, t)$ (recall that these potentials are independent of $k$). Consequently, due to isotropic and scale-invariant initial conditions, and scale-invariant (but not isotropic) dynamics, the anisotropies acquired by long-wavelength perturbations are scale-invariant. Stated in a different way, given the point (ii) above, the dynamics for the CMB modes of interest during the time anisotropies are non-negligible are scale-invariant, namely independent of $k$. Combined with the first fact that the initial conditions $P^{LM}_{\cal R}(k) = 0$ for $L \neq 0$, it follows that the anisotropies will generate non-zero $P^{LM}_{\cal R}(k)$ for $L \neq 0$, essentially sourced by $P^{00}_{\cal R}(k)$ and therefore all $P^{LM}_{\cal R}(k)$ will have exactly the same scale-dependence as $P^{00}_{\cal R}(k)$: a slight red tilt. Then, when $P^{00}_{\cal R} (k)$ is factored out to define the $g_{LM}(k)$ as is done in \eqref{anisPg}, the identical small red tilt in each $P^{LM}_{\cal R}(k)$ is entirely captured by the $P^{00}_{\cal R} (k)$ prefactor, with the result that all $g_{LM}(k)$ are exactly scale-invariant. As a result, anisotropies in these bouncing cosmologies are dominated by a scale-invariant quadrupolar contribution. (As an aside, note that this would not have been the case if anisotropies were large at the time the observable modes exited the Hubble radius, but this possibility would require much larger anisotropies in the background and is ruled out by observations. In the same way, shorter-wavelength modes in these models that are super-Hubble for only a portion of the time anisotropies are significant will not have a $g_2$ that is scale-invariant, but these modes are not relevant for the CMB.)

This scale invariance has an important consequence: since $g_2$ is independent of $k$, it is impossible to `erase' anisotropies in the perturbations during the post-bounce expansion by red-shifting modes with large $g_2$ to super-Hubble scales today. This is an important difference with anisotropies in the background space-time, which scale as $\sigma^2 \propto a^{-6}$ and therefore are rapidly diluted in an expanding universe. In other words, the expansion of the universe isotropizes the homogeneous metric, but not the perturbations---this is the reason observational constraints are so strong.

The scale-invariance is also an important difference with inflationary cosmology. Anisotropies can in principle be important during the early stages of inflation, and generate anisotropies in the CMB as well; however, since anisotropies are only imprinted on super-horizon perturbations and anisotropies rapidly decay in an inflationary universe, the predicted scale-dependence for $g_2$ is strongly red, with an approximate scaling of $g_2 \sim k^{-1}$ \cite{Agullo:2020wur}.

Therefore, a hallmark of bouncing alternatives to inflation is a scale-invariant anisotropy in the CMB with the dominant contribution being the quadrupolar moment $g_2$; this scale-invariance offers a potential observational test that could differentiate these models from inflationary cosmology. In Sec.~\ref{pred}, we also compute the predictions for angular correlation functions.

\section{Observational constraints on anisotropies}
\label{upp}

In this section, we revisit the same type of calculations we presented in the previous section, but from a different perspective and with a few changes to the initial conditions.

In Sec.~\ref{dyn}, we compared numerical solutions for background geometries with different matter fields that have the same initial conditions for both the energy density and for the shears. Here, we will allow the initial conditions for the shears to vary, selecting its initial value $\sigma_{in}^2$ so that the predicted $g_2$ is comparable to the observational limit \eqref{g2-planck}---of course, this means that $\sigma_{in}^2$ will vary for different matter fields. This calculation will inform us about the maximum value of ${}^{(r)}\sigma^{2}_{in,max}$, ${}^{(s)}\sigma^{2}_{in,max}$ and ${}^{(e)}\sigma^{2}_{in,max}$, namely, the maximum value that the shear squared can take at the time when the energy density equals $\rho_{in}=4.96 \times 10^{-18}$ for a contracting universe dominated by radiation, stiff matter, or an ekpyrotic fluid, respectively. (The maximum value for the shear at other values of $\rho$ can be calculated using their respective scalings, $\sigma^2 \sim a^{-6}$ and $\rho \sim a^{-3(1+w)}$.)

\begin{figure}[t]
{\centering
\includegraphics[width = 0.48\textwidth]{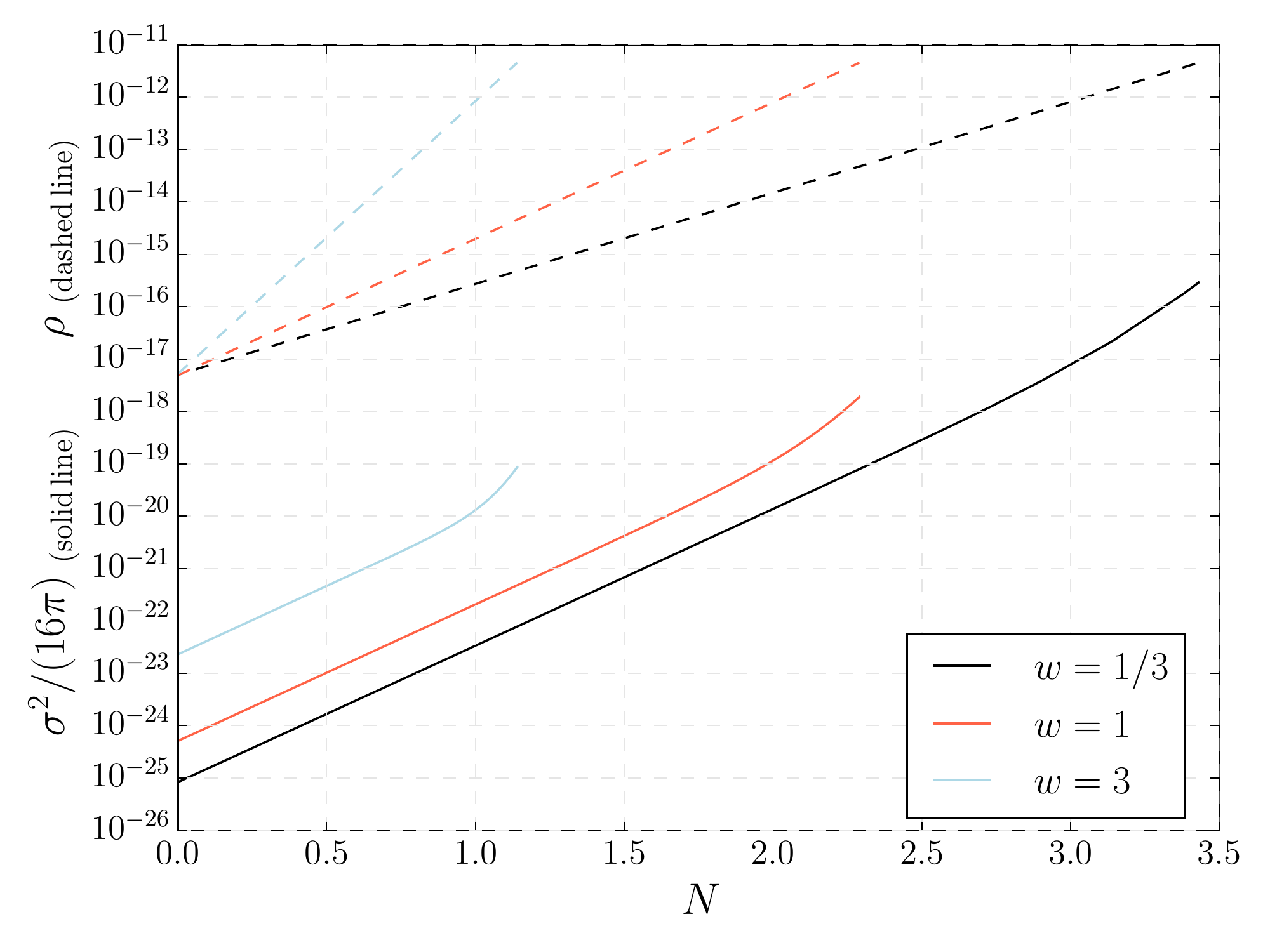}
\includegraphics[width = 0.48\textwidth]{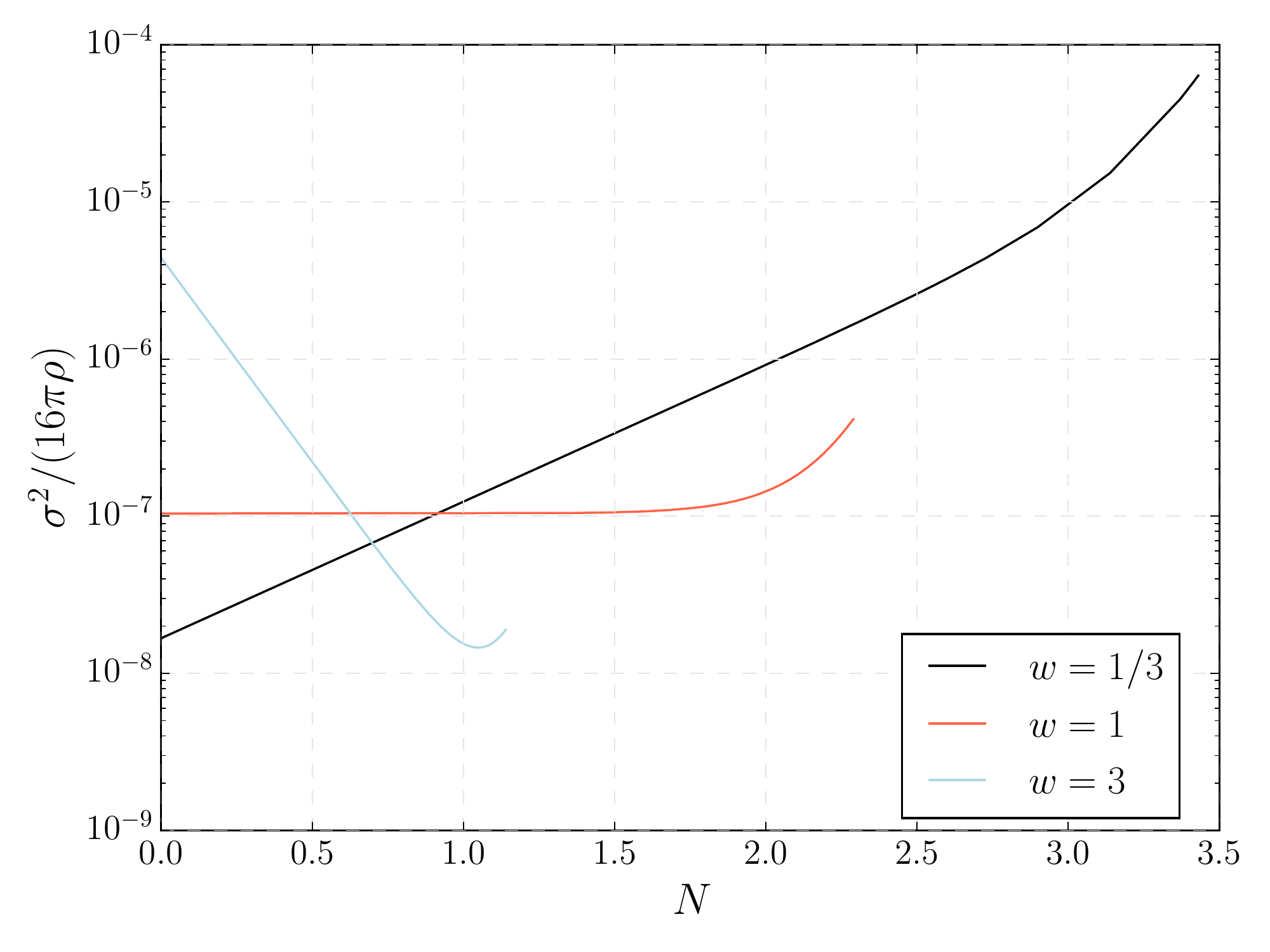}
}
\caption{These plots show the numerical solutions for the background space-time with initial conditions that saturate the observational upper bound for $g_2$, given the same initial energy density $\rho_{in}$ as for Fig.~\ref{fig2}, and the same $\rho_b$. As in Fig.~\ref{fig2}, the total number of $e$-folds to the bounce is $N_e=1.14$, $N_s=2.29$ and $N_r=3.43$, for ekpyrosis, stiff matter, and radiation, respectively. \\
\textit{Left panel:} This plot shows $\rho$ (dashed lines) and $\sigma^2/(16\pi)$ (solid lines) as a function of the number of $e$-folds of contraction from the initial time to the bounce. The initial values of $\rho$ are the same as in Fig.~\ref{fig2}, while those of $\sigma^2/(16\pi)$ are ${}^{(r)}\sigma^2_{in,max} / (16\pi) = 8.36 \times 10^{-26}$, ${}^{(s)}\sigma^2_{in,max} / (16\pi) = 5.10 \times 10^{-25}$, and ${}^{(e)}\sigma^2_{in,max} / (16\pi) = 2.30 \times 10^{-23}$, while the maximum values of $\sigma^2/(16\pi)$ are reached at the bounce, and are ${}^{(r)}\sigma^2_{b,max} / (16\pi) = 2.90 \times 10^{-16}$, ${}^{(s)}\sigma^2_{b,max} / (16\pi) = 1.89 \times 10^{-18}$ and ${}^{(e)}\sigma^2_{b,max} / (16\pi) = 8.61 \times 10^{-20}$, all in Planck units. \\
\textit{Right panel:} This plot shows the ratio $\sigma^2/(16\pi\rho)$ as a function of the number of $e$-folds of contraction from the initial time to the bounce. For radiation this ratio grows monotonically, reaching the maximum value ${}^{(r)}\sigma^2_r / (16\pi\rho_{b,max}) = 6.37 \times 10^{-5}$ at the bounce, while for stiff matter the ratio is constant except near the bounce where it grows slightly (due to departures from general relativity) to ${}^{(s)}\sigma^2_{b,max} / (16\pi\rho_b) = 4.16 \times 10^{-7}$. This ratio decreases in the ekpyrotic scenario to a minimum value of ${}^{(e)}\sigma^2_{min} / (16\pi\rho_{min}) = 1.46 \times 10^{-8}$ at $N=1.04$, and then slightly grows just before the bounce (again, due to departures from general relativity near the bounce).}
\label{fig3}
\end{figure}

The results of the simulations are presented in Fig.~\ref{fig3}, which are analogous to the plots shown in Fig.~\ref{fig2} except with different initial conditions for the shears (as explained above). For the runs presented here, the predicted value for $g_2$ is smaller than the observational bound \eqref{g2-planck}, but of the same order of magnitude---specifically for the radiation run $g_2 = 6.00 \times 10^{-3}$, while for the stiff matter run $g_2 = 7.50 \times 10^{-3}$, and for the ekpyrotic ($w=3$) case, $g_2 = 7.11 \times 10^{-3}$. The main features we find are the following.

First, since both the value of the energy density at the initial time of our simulations, $\rho_{in}$, as well as its value at the bounce, $\rho_b$, are unchanged, the energy density (shown as dashed lines in the left panel of Fig.~\ref{fig3}) evolves in exactly the same way as in the previous section.

On the other hand, the initial conditions for $\sigma^2$ have changed, and the solid lines in the left panel of Fig.~\ref{fig3} show two important results. First, observational constraints give different bounds on the possible initial values for the shear, with
\be
{}^{(r)}\sigma^{2}_{in,max}\, <\, {}^{(s)}\sigma^{2}_{in,max}\, < \, {}^{(e)}\sigma^{2}_{in,max}\, ,
\ee
so the initial value of $\sigma^{2}$ can be larger if the subsequent evolution is dominated by an ekpyrotic fluid. This result may not be surprising, but this relation is inverted closer to the bounce where
\be
{}^{(r)}\sigma^{2}_{b,max}\, >\, {}^{(s)}\sigma^{2}_{b,max}\, > \, {}^{(e)}\sigma^{2}_{b,max}\, ,
\ee
where the subscript $b$ denotes the bounce. In other words, given observational constraints on $g_2$, shear anisotropies at the bounce can be larger in a radiation-dominated universe than in ekpyrosis. Note that it is not the maximal value of $\sigma^2$ or $\sigma^2/\rho$ that is constrained; the relation is richer and more complex than this, since anisotropies in perturbations are acquired continuously throughout the evolution. (It is tempting to look for other simple measures of anisotropies that could be used, for example $\int \dd t ~ \sigma^2$ or $\int \dd t ~ \sigma^2/ (16 \pi \rho)$, but these and other relatively simple measures fail to capture the complexity of the dynamics for perturbations in an anisotropic background.)

Also, the right panel shows the ratio $\sigma^2/(16\pi\, \rho)$ for each of the three runs. Keeping in mind that these runs have the maximal anisotropies allowed (for the matter fields we consider) given CMB data, note that the observational constraints are much stronger than simply requiring $\sigma^2/(16\pi\, \rho) \ll 1$. Instead, CMB data implies for these runs that $\sigma^2/(16\pi\, \rho) < 10^{-4}$ at all times, and the constraints are even stronger for ekpyrosis and stiff matter. Clearly, perturbations (combined with observations) provide a powerful probe of anisotropies in bouncing universes.

Finally, it is also interesting to compute the maximum allowed value of $\sigma^2/(16\pi\, \rho)$ at the time of horizon exit for the primordial perturbations in the contracting phase. This calculation requires an assumption about the dominant matter content of the universe from horizon exit to the time at which we start our simulations, $t_{in}$. For completeness, we report here the result for two possibilities, matter-dominated and radiation-dominated contraction. Let us focus on the reference scale $k_*$ whose physical value today is $0.05\, {\rm Mpc}^{-1}$, and denote by $t_{\rm ext}$ the time at which it exited the Hubble radius in the contracting phase. For a matter-dominated universe between $t_{\rm ext}$ to $t_{in}$, there must be 120 $e$-folds of contraction between these two times and, consequently, the maximum value of $\sigma^2/(16\pi\, \rho)$ at $t_{\rm ext}$ is about 157 orders of magnitude smaller than at $t_{in}$, in general agreement with earlier estimates \cite{Levy:2016xcl}. If the universe were dominated by radiation between these two instants, then there would be 60 $e$-folds of expansion in between, and the maximum value of $\sigma^2/(16\pi\, \rho)$ at $t_{\rm ext}$ would be about 53 orders of magnitude smaller than at $t_{in}$. These numbers slightly overestimate the upper bound for $\sigma^2/(16\pi\, \rho)$ at $t_{\rm ext}$, since in obtaining them we have neglected the anisotropies induced in the perturbations during the evolution between $t_{\rm ext}$ and $t_{in}$. These anisotropies should be very small though, since the values of $\sigma^2(t)$ are constrained to be extraordinarily small.

\section{Predictions for the CMB: angular correlation functions}
\label{pred}

We found in the previous section that the main signature of anisotropies in bouncing alternatives to inflation is a scale-invariant quadrupolar modulation in the primordial scalar power spectrum; subleading effects are higher-order modulations, as well as cross-correlations between scalar and tensor modes. For completeness, in this section we compute some of the imprints these anisotropies leave in the angular correlation functions in the CMB. We present results for the background radiation-dominated Bianchi~I space-time with maximal anisotropies compatible with the observational constraints, with the initial conditions for the background and for the perturbations as given in Sec.~\ref{upp}, for which $g_2 \sim g_2^{\rm obs} = 7.62 \times 10^{-3}$. (For the radiation-dominated case, the same initial conditions are used in both of Secs.~\ref{dyn} and \ref{upp}.)

Specifically, we set the initial data for the background as given in Table~\ref{indata} for a radiation-dominated universe at $t=t_{in}$, and assume that scalar perturbations are initially in a state with a power spectrum of the form of Eq.~\eqref{init-P}, with $A_{in} = 8.78 \times 10^{-17}$ and $k_{ref} = 1.29 \times 10^{-33}$ (in Planck units).

Recall that, as discussed in the previous section, the amplitude of tensor perturbations generated before the bounce is model dependent (with ekpyrosis not producing significant tensor modes, while the matter bounce typically produces a large amplitude of tensor modes). For this reason, we have focused so far on scalar perturbations, and set initial conditions for the tensor modes to be $\Gamma_1(t_{in}) = \Gamma_2(t_{in}) = 0$. We will use the same initial conditions for tensor modes in this section as well, but even with these initial conditions, it is illustrative to show the angular correlation functions for tensor modes as well, since tensor modes are excited dynamically through a coupling to scalar perturbations in anisotropic space-times. For the background radiation-dominated Bianchi~I space-time given in Sec.~\ref{upp}, this coupling generates scale-invariant tensor modes with a tensor-to-scalar ratio $r = 5.18 \times 10^{-7}$. (Of course, if tensor modes are excited during the contraction phase before anisotropies become important, as is the case for the matter bounce scenario, then the tensor-to-scalar ratio can be much larger than this, although it must respect the observational constraint $r < 0.036$ \cite{BICEPKeck:2021gln}.)

The angular correlation functions of temperature anisotropies $T(\hat n)$ and the electric and magnetic components of the polarization fields of CMB photons, $E(\hat n)$ and $B(\hat n)$, are defined as
\be 
C_{\ell \ell^{\prime},m m^{\prime}}^{X, X^{\prime}}=\left\langle a_{\ell m}^{X} a_{\ell^{\prime} m^{\prime}}^{X^{\prime} }\right\rangle,
\ee
where $a^X_{\ell m}$ are the angular-Fourier components of the fields $X(\hat n)=\{T(\hat n), E(\hat n), B(\hat n)\}$,
\be 
a^X_{\ell m} = \int \dd \Omega\, X(\hat n)\, \bar Y_{\ell m}(\hat n)\, ,\ \ X=\{T,E,B\} ,
\ee
and $Y_{\ell m}(\hat n)$ are the usual (scalar) spherical harmonics (see Sec.~V in \cite{Agullo:2020iqv} for details omitted here). The amplitudes $a^X_{\ell m}$ can be obtained from the primordial scalar and tensor perturbations as
\be \label{alm}
a^X_{\ell m}=\int \frac{d^3 k}{(2\pi)^3}\, (-i)^{\ell}\, \sum_{s=0,\pm2} {_s\Delta}^X_{\ell}(k)\, \Gamma_s(\vec k)\; {_s\bar{Y}}_{\ell m}(\hat k)\, ,
\ee
where ${_s\bar{Y}}_{\ell m}(\hat k)$ denote spherical harmonics with spin-weight $s$. For convenience, we have changed the basis in the tensor perturbations from $\Gamma_1$ and $\Gamma_2$ associated with the $+$ and $\times$ polarizations to $\Gamma_{\pm 2}$ associated with circular polarization, with
\be
\Gamma_{\pm 2}(\vec k)=\frac{1}{\sqrt{2}}\left(\Gamma_1(\vec k) \mp i\, \Gamma_2(\vec k)\right)\, .
\ee
We use the sub-index $s=0,\pm 2$ rather than $\mu=0,1,2$ to emphasize that we work with circularly polarized tensor modes. 

The functions $_s\Delta^X_{\ell}(k)$ in Eq.~\eqref{alm} are the transfer functions that encode the complex physics involved in the process of evolving the fields $\Gamma_s$ across the radiation dominated era from the time observable modes re-enter the Hubble radius to the formation of the CMB. They can be obtained, for instance, by using a Boltzmann code such as CLASS \cite{Lesgourgues:2011re}. Note from Eq.~\eqref{alm} that each amplitude $a^X_{\ell m}$ can be sourced from the three primordial perturbations $\Gamma_0$, $\Gamma_{\pm 2}$. The scalar perturbations $\Gamma_0$ do not contribute to the $B$-polarization in the CMB because $_0\Delta^B_{\ell}(k)=0$. 

Also note that  we are using isotropic transfer functions here (for anisotropic transfer functions see \cite{Saadeh:2016sak}). The transfer functions used to propagate linear perturbations depend on the background metric, but not on the state of perturbations themselves, due to the absence of backreaction in the approximation of linear perturbation theory. Since the expansion isotropizes the metric, for the space-times considered here, at the time observable modes re-enter the horizon the shears are extraordinarily small, and the use of isotropic transfer functions is an excellent approximation.

Using Eq.~\eqref{alm}, the angular correlation functions $C_{\ell \ell^{\prime},m m^{\prime}}^{X, X^{\prime}}$ can be written in terms of the primordial power spectra ${{\cal P}}_{ss'}(\vec{k})$,
\be 
C_{\ell \ell^{\prime}, m m^{\prime}}^{X, X^{\prime}}=\int \frac{d^3 k}{(2\pi)^3}\, (-i)^{(\ell+\ell')} \, \sum_{s,s'} \, {_s\Delta}^X_{\ell}(k)\, {_{s'}\Delta}^{X'}_{\ell'}(k)\,\frac{2\pi^2}{k^3}{{\cal P}\,}_{ss'}(\vec{k})\, {_s\bar{Y}}_{\ell m}(\hat k)\, {_{s'}\bar{Y}}_{\ell' m'}(-\hat k)\, . 
\ee
The parity invariance of the quantum state describing the perturbations implies that $ C_{\ell \ell^{\prime}, m m^{\prime}}^{X, X^{\prime}}$ are also parity-invariant, and therefore
\bea
C_{\ell \ell^{\prime}, m m^{\prime}}^{TT}&=&C_{\ell \ell^{\prime}, m m^{\prime}}^{EE}=C_{\ell \ell^{\prime}, m m^{\prime}}^{BB}=C_{\ell \ell^{\prime}, m m^{\prime}}^{TE}=0\, \ \ \ \ \ {\rm if} \ \ell+\ell^{\prime}\, {\rm odd}\, , \\ 
C_{\ell \ell^{\prime}, m m^{\prime}}^{TB}&=& C_{\ell \ell^{\prime}, m m^{\prime}}^{EB}=0\, \ \ \ \ \ {\rm if} \ \ell+\ell^{\prime}\, {\rm even} \, .
\eea
For cosmological perturbations on FLRW space-times, isotropy would further impose the condition that all angular correlation functions vanish unless $\ell=\ell'$. But in anisotropic Bianchi~I spacetimes, the cross-correlations $C_{\ell \ell^{\prime}, m m^{\prime}}^{TB}$ and $C_{\ell \ell^{\prime}, m m^{\prime}}^{EB}$ can be non-zero, although only when the combination $\ell+\ell'$ is odd. Correlations of this type are therefore a smoking gun signature for anisotropies in the early universe.

Since anisotropies are small, it is to be expected that the largest correlation functions will be T-T, then E-T and E-E, all for the case $\ell=\ell'$. Smaller correlations are found for B-B, as well as off-diagonal $\ell \neq \ell'$ correlations, which are non-vanishing in contrast to the isotropic case. In particular, T-B and E-B off-diagonal cross-correlations are non-vanishing for odd $\ell+\ell'$.

\begin{figure}[th]
{\centering
\includegraphics[width = 0.49\textwidth]{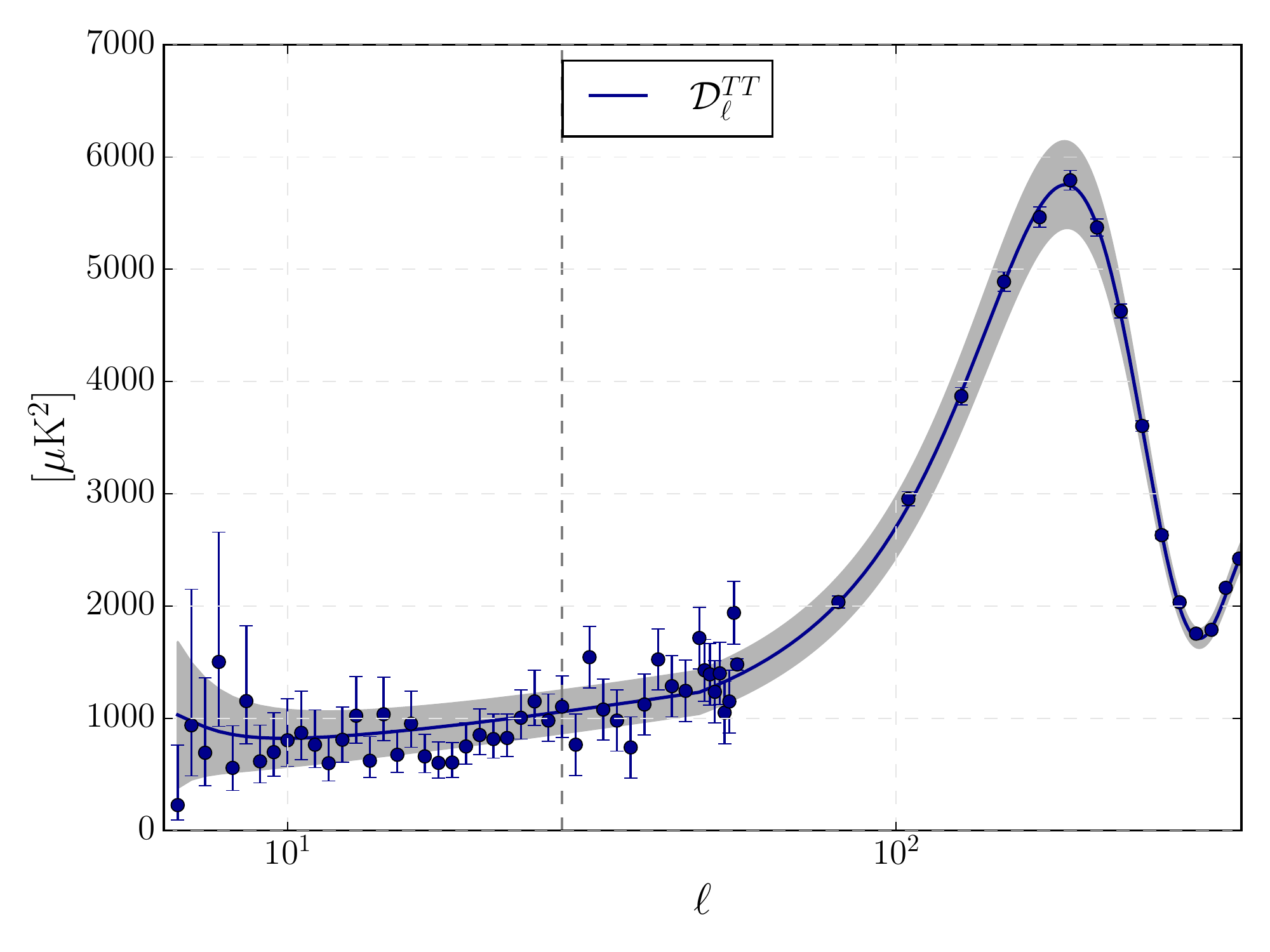}
\includegraphics[width = 0.49\textwidth]{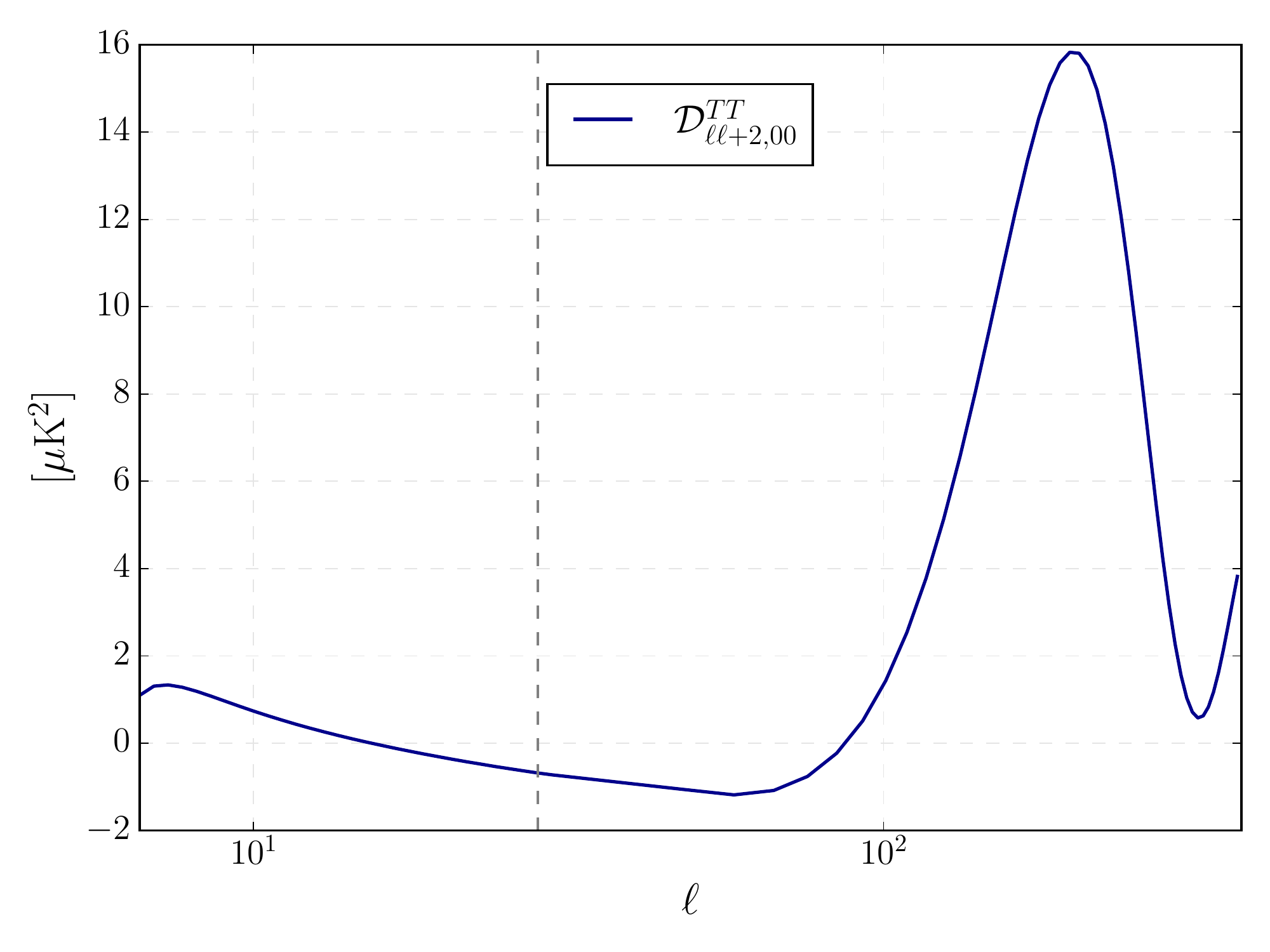}
}
\caption{T-T angular correlation functions for perturbations on a radiation-dominated Bianchi~I background with the initial conditions described in Sec.~\ref{upp}. The vertical dashed line denotes the transition from a linear scale for $\ell$ (on its left) to a logarithmic scale (on its right). \\
\textit{Left panel:} The temperature-temperature angular correlation function $D_{\ell}^{TT}$ is shown by the solid line, while the dots (and uncertainties) are the observations from the Planck collaboration \cite{Planck-data}, with the shaded region showing the cosmic variance uncertainty. \\
\textit{Right panel:} The temperature-temperature angular correlation function $D_{\ell \ell+2, 00}^{TT}$. This correlation function would be identically zero in the absence of primordial anisotropies.}
\label{fig:DTT}
\end{figure}

First, the T-T angular correlation function is shown in Fig.~\ref{fig:DTT}, with the left panel showing the diagonal $\ell=\ell'$ case, specifically the quantity
\be
D_{\ell}^{TT} = \frac{T_0^2 \ell (\ell+1)}{2\pi}\, C_{\ell}^{TT}\, ,
\ee
where $T_0$ is the CMB average temperature and $C_{\ell}^{TT}=\frac{1}{2\ell+1} \sum_{m=-\ell}^\ell (-1)^mC_{\ell \ell, m -m}^{TT}$. Although all primordial power spectra ${{\cal P}}_{ss'}(\vec k)$, with $s,s'=0,\pm2$, contribute to $C_{\ell \ell^{\prime}, m m^{\prime}}^{TT}$, the scalar one $s=0=s'$ gives the dominant contribution. We also show Planck's observations \cite{Planck-data} for comparison as data points in the left panel; since the differences are smaller than cosmic variance (indicated by the shaded region) this correlation function cannot be used to detect small primordial anisotropies; rather it is necessary to consider other correlation functions.

One possible example of a correlation function that is sensitive to small primordial anisotropies is
the first non-trivial off-diagonal T-T correlation function for $\ell = \ell'+2$. The right panel of Fig.~\ref{fig:DTT} shows the quantity
\be
D_{\ell \ell+2, 00}^{TT} = \frac{T_0^2 \ell (\ell+1)}{2\pi}\, C_{\ell \ell+2, 00}^{TT}\, .
\ee
This correlation function would be identically zero in FLRW spacetimes and consequently it is a powerful probe for primordial anisotropies. It is also possible to compute other angular correlation functions, for instance with $\ell'=\ell+4$, $\ell'=\ell+6$, etc., or different values of $m$ or $m'$. But $D_{\ell \ell+2, 00}^{TT}$ is a good representative example of the anisotropic correlation functions expected in bouncing models, which are dominated by a quadrupolar angular distribution. Note that since this observable (and others like it defined below) selects the specific values of $m=m'=0$, it is not invariant under coordinate rotations and its predicted spectrum is understood to be for the set of coordinates for the Bianchi~I background spacetime where $\Psi=0$.

\begin{figure}[th]
{\centering
\includegraphics[width = 0.49\textwidth]{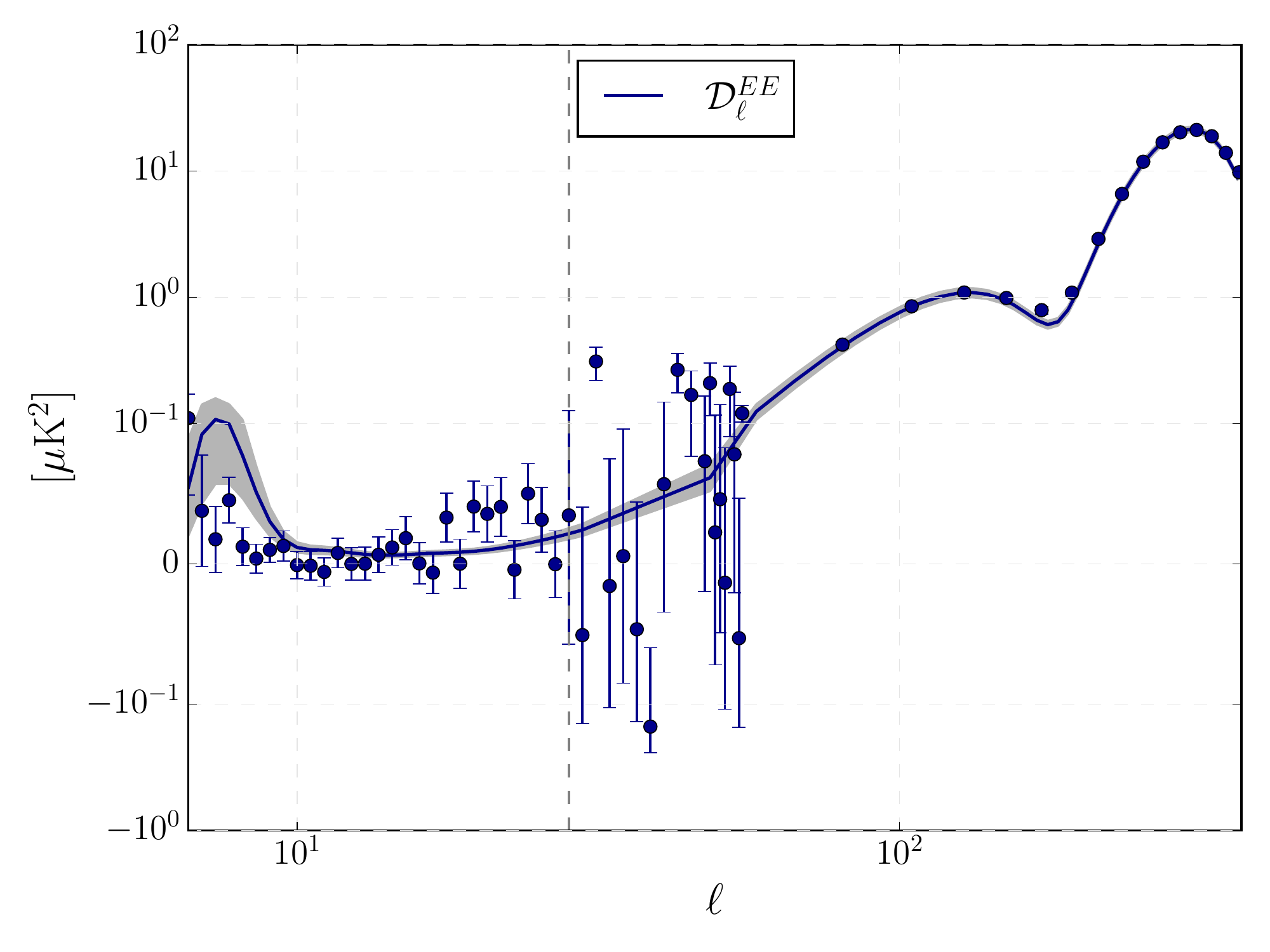}
\includegraphics[width = 0.49\textwidth]{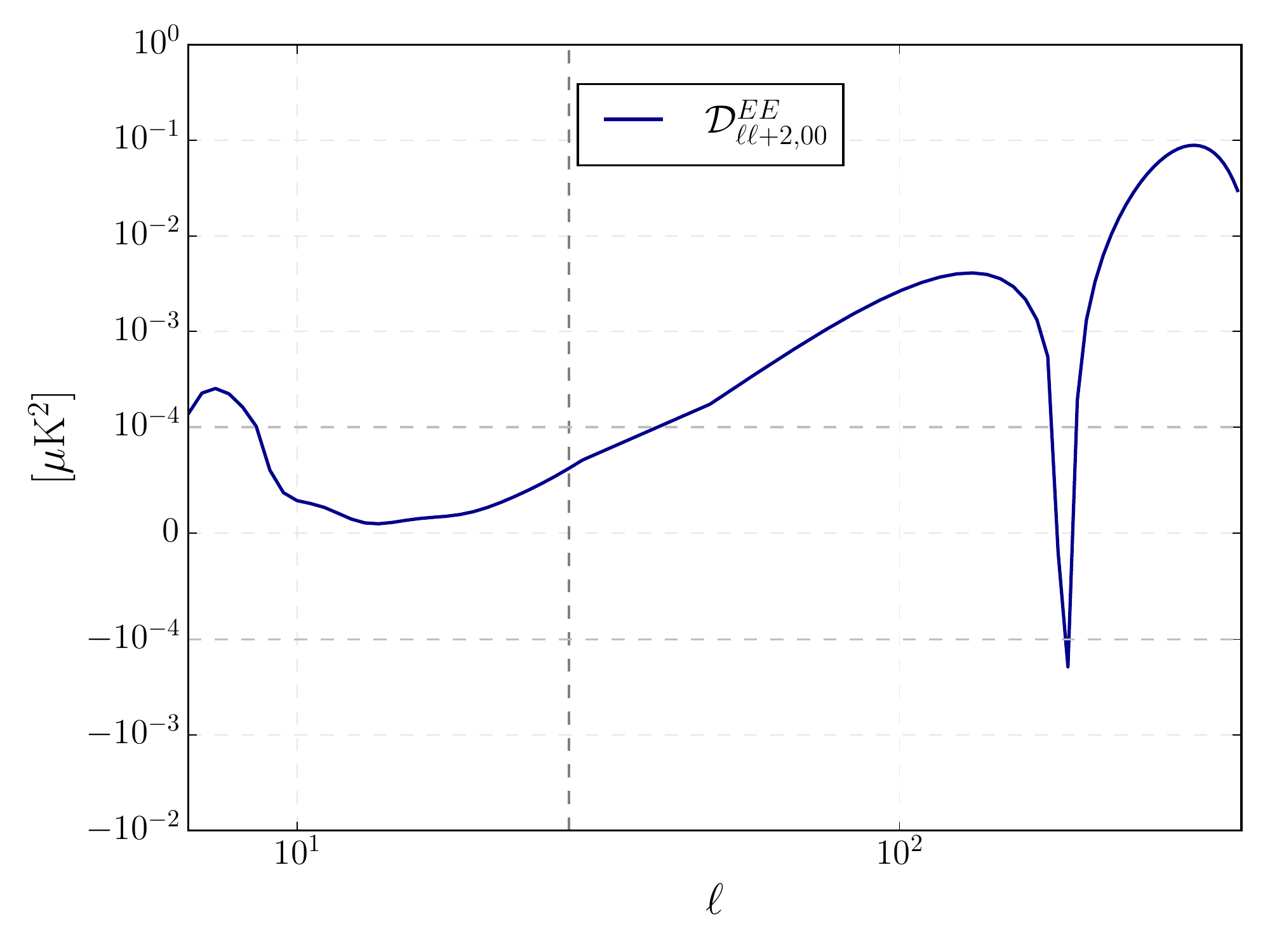}
}
\caption{E-E angular correlation functions for perturbations on a radiation-dominated Bianchi~I background with the initial conditions described in Sec.~\ref{upp}. The vertical dashed line denotes the transition from a linear scale for $\ell$ (on its left) to a logarithmic scale (on its right). \\
\textit{Left panel:} The E-E angular correlation function $D_{\ell}^{EE}$ is shown by the solid line, while the dots (and uncertainties) are the observations from the Planck collaboration \cite{Planck-data}, and the shaded region shows the cosmic variance uncertainty. \\
\textit{Right panel:} The E-E angular correlation function $D_{\ell \ell+2, 00}^{EE}$. This correlation function would be identically zero in the absence of primordial anisotropies.}
\label{fig:DEE}
\end{figure}

Second, we calculate E-E correlation functions. The largest correlations are for the $\ell=\ell'$ case, and the quantity
\be
D_{\ell}^{EE} = \frac{T_0^2 \ell (\ell+1)}{2\pi}\, C_{\ell}^{EE}\, ,
\ee
where $C_{\ell}^{EE}=\frac{1}{2\ell+1} \sum_{m=-\ell}^\ell (-1)^mC_{\ell \ell, m -m}^{EE}$, is plotted in the left panel of Fig.~\ref{fig:DEE}, and compared with the Planck Collaboration's data points \cite{Planck-data}, with the shaded region representing the uncertainty due to cosmic variance. Similarly to the T-T correlations, the effect of primordial anisotropies is negligibly small in $D^{EE}_{\ell \ell}$.

In the right panel of Fig.~\ref{fig:DEE}, we show the anisotropic correlation function
\be
D_{\ell \ell+2, 00}^{EE} = \frac{T_0^2 \ell (\ell+1)}{2\pi}\, C_{\ell \ell+2, 00}^{EE}\, ,
\ee
which vanishes in the isotropic case, while here it takes values of about $0.1\%$ with respect to $D_{\ell}^{EE}$. This correlation function is given as a representative example; other angular correlation functions with $\ell'=\ell+4$, $\ell'=\ell+6$, etc., and different values of $m$ or $m'$ are also non-zero.

\begin{figure}[th]
{\centering
\includegraphics[width = 0.49\textwidth]{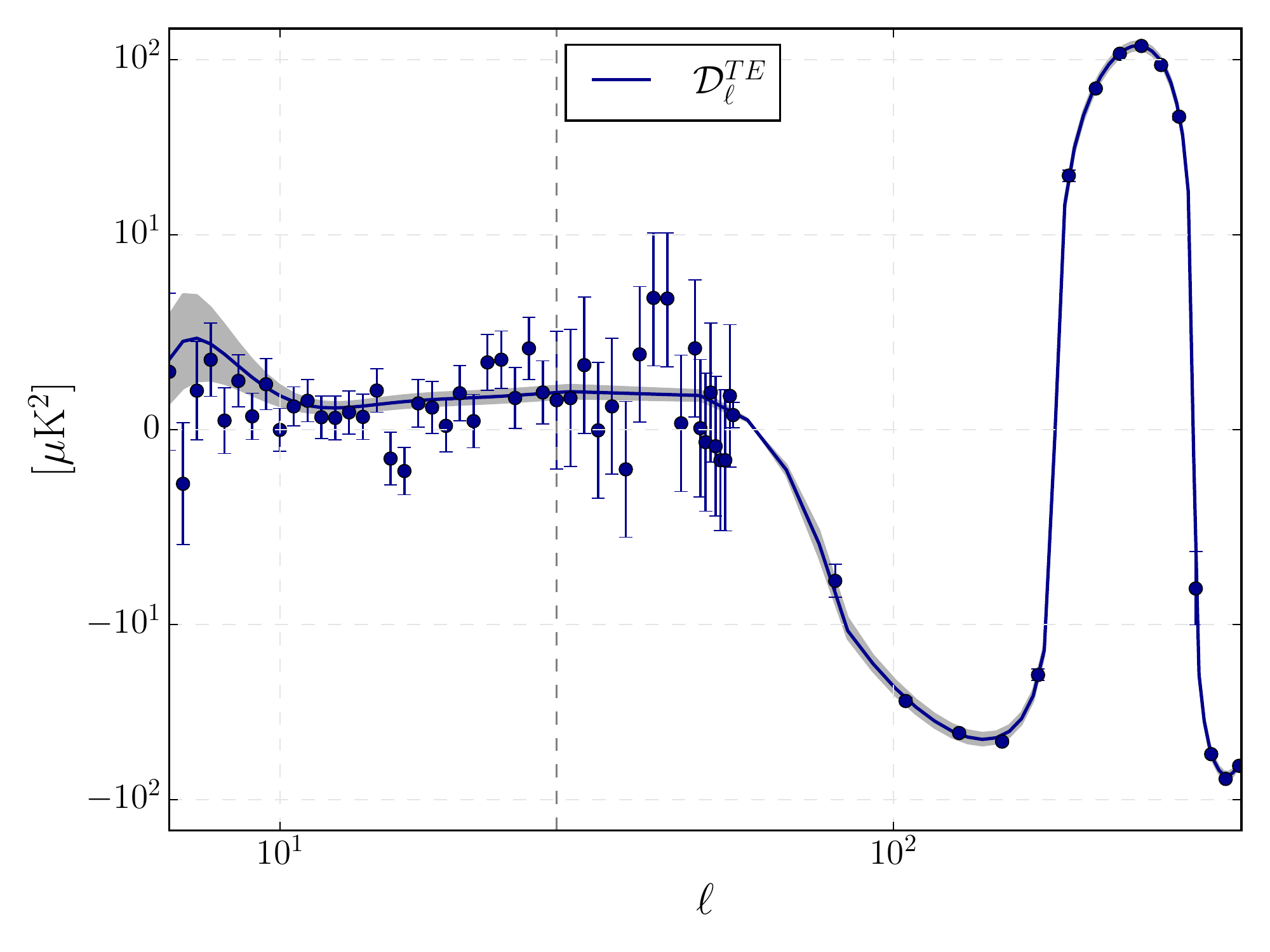}
\includegraphics[width = 0.49\textwidth]{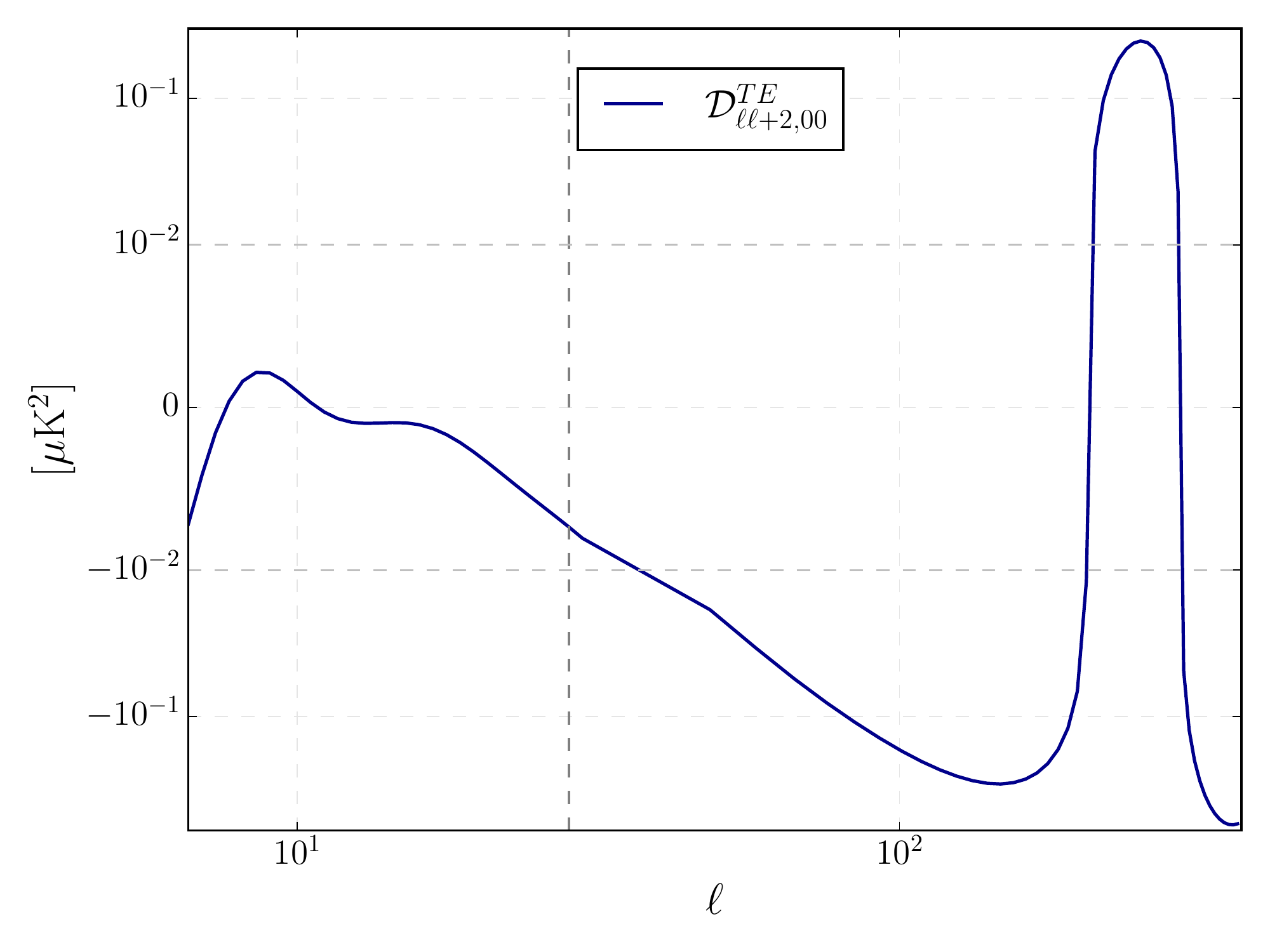}
}
\caption{T-E angular cross-correlation functions for perturbations on a radiation-dominated Bianchi~I background with the initial conditions described in Sec.~\ref{upp}. The vertical dashed line denotes the transition from a linear scale for $\ell$ (on its left) to a logarithmic scale (on its right). \\
\textit{Left panel:} The T-E angular correlation function $D_{\ell}^{TE}$ is shown by the solid line, while the dots (and uncertainties) are the observations from the Planck collaboration \cite{Planck-data}, and the shaded region shows the cosmic variance uncertainty. \\
\textit{Right panel:} The T-E angular correlation function $D_{\ell \ell+2, 00}^{TE}$. This correlation function would be identically zero in absence of primordial anisotropies.}
\label{fig:DTE}
\end{figure}

Third, T-E cross-correlation functions are shown in Fig.~\ref{fig:DTE}. The left panel shows the diagonal angular correlation function
\be
D_{\ell}^{TE} = \frac{T_0^2 \ell (\ell+1)}{2\pi}\, C_{\ell}^{TE}\, ,
\ee
where $C_{\ell}^{TE}=\frac{1}{2\ell+1} \sum_{m=-\ell}^\ell (-1)^mC_{\ell \ell, m -m}^{TE}$, together with the data points from the Planck Collaboration \cite{Planck-data}, and the shaded region shows the uncertainty due to cosmic variance. Additionally, the right panel contains a plot of the off-diagonal angular correlation function
\be
D_{\ell \ell+2, 00}^{TE}\equiv \frac{T_0^2 \ell (\ell+1)}{2\pi}\, C_{\ell \ell+2, 00}^{TE}\, ,
\ee
that vanishes in the isotropic case. As in the previous angular correlation functions, other off-diagonal components of these correlation functions with $\ell'=\ell+4$, $\ell'=\ell+6$, etc., and different values of $m$ or $m'$ are also non-zero.

\begin{figure}[th]
{\centering
\includegraphics[width = 0.49\textwidth]{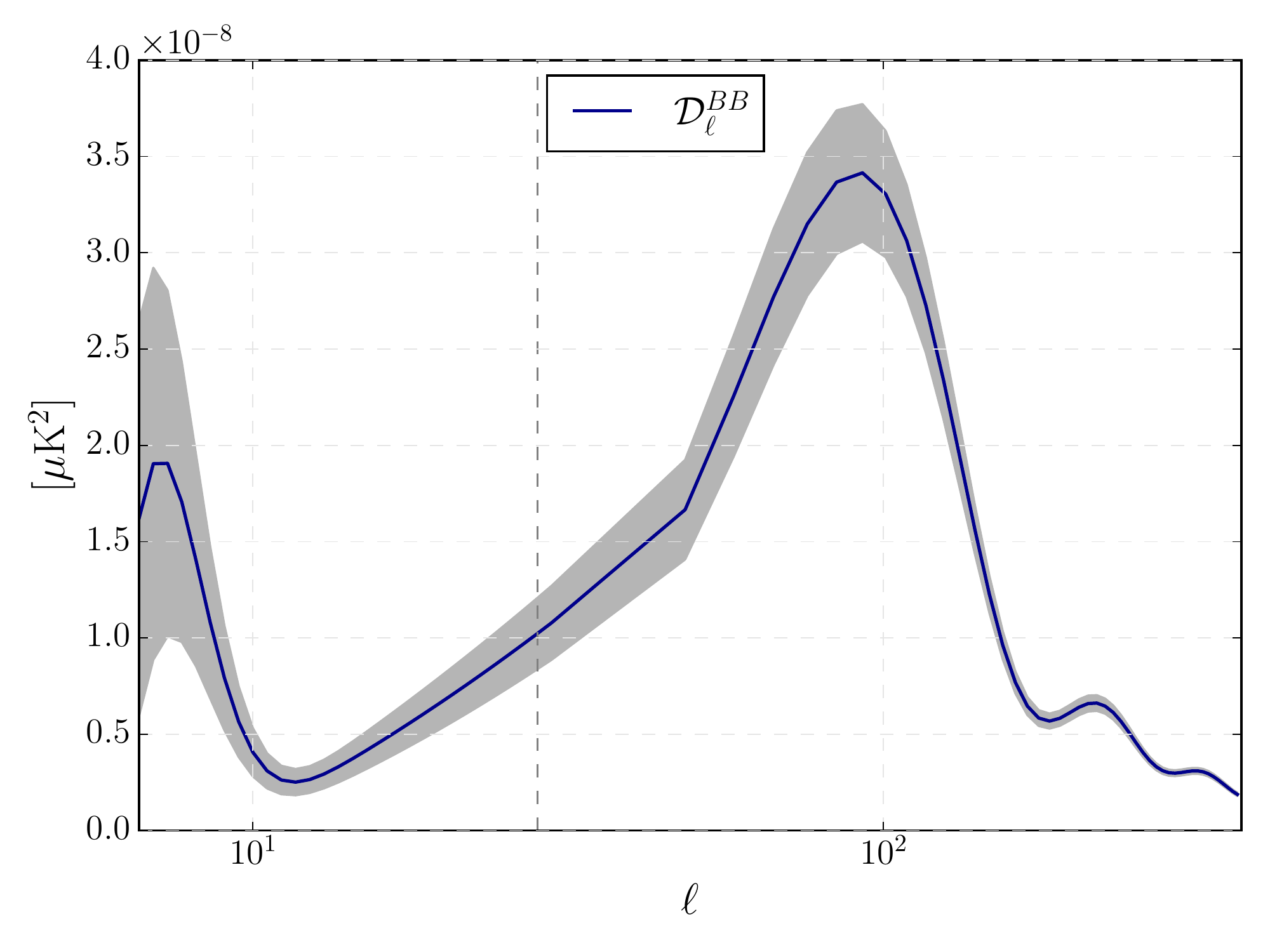}
\includegraphics[width = 0.49\textwidth]{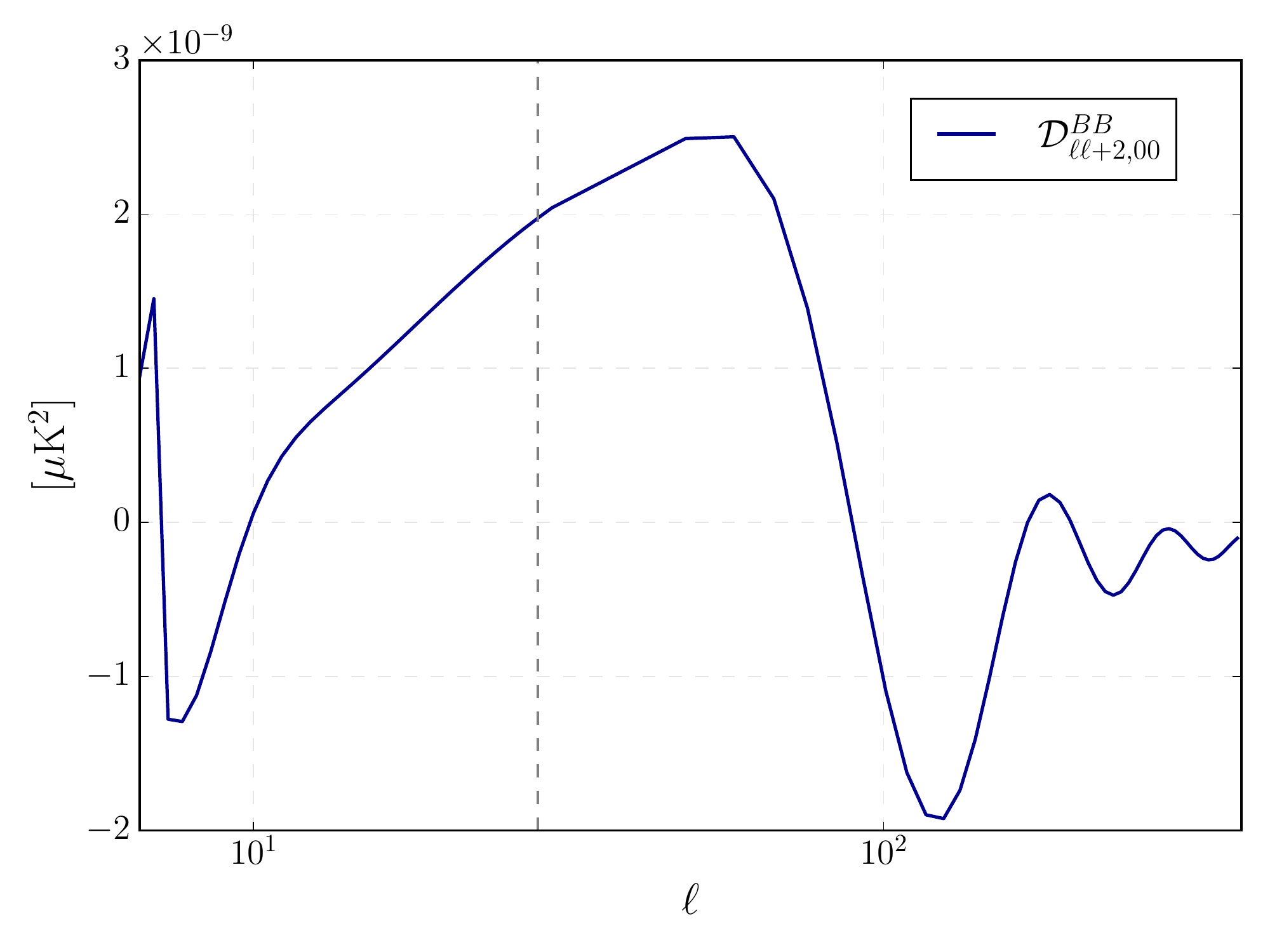}
}
\caption{B-B angular correlation functions for perturbations on a radiation-dominated Bianchi~I background with the initial conditions described in Sec.~\ref{upp}. The vertical dashed line denotes the transition from a linear scale for $\ell$ (on its left) to a logarithmic scale (on its right). \\
\textit{Left panel:} The B-B angular correlation function $D_{\ell}^{BB}$ is shown by the solid line, while the shaded region shows the cosmic variance uncertainty. This gives a tensor-to-scalar ratio of $r \sim 5 \times 10^{-7}$. \\
\textit{Right panel:} The B-B angular correlation function $D_{\ell \ell+2, 00}^{BB}$. This correlation function would be identically zero in absence of primordial anisotropies.}
\label{fig:DBB}
\end{figure}

Fourth, B-B correlation functions are shown in Fig.~\ref{fig:DBB}. Since only the purely tensorial primordial spectra ${{\cal P}}_{\pm 2\pm 2}(\vec k)$ and ${{\cal P}}_{\mp 2\pm 2}(\vec k)$ contribute to these correlations, the result will clearly depend on the initial pre-bounce amplitude of the tensor modes. Here we present results for the case when the tensor modes are initially vanishing (this initial condition is most relevant for ekpyrotic models), but these calculations can easily be extended to the case where primordial tensor modes are generated during the contracting phase (for example in a matter bounce scenario).

The left panel of Fig.~\ref{fig:DBB} shows
\be
D_{\ell}^{BB}\equiv \frac{T_0^2 \ell (\ell+1)}{2\pi}\, C_{\ell}^{BB}\, ,
\ee
where $C_{\ell}^{BB}=\frac{1}{2\ell+1} \sum_{m=-\ell}^\ell (-1)^mC_{\ell \ell, m -m}^{BB}$, together with the uncertainty due to cosmic variance. As stated above, this corresponds to a tensor-to-scalar ratio of $r \sim 5 \times 10^{-7}$. The right panel shows one of the off-diagonal elements of this correlation function, specifically
\be
D_{\ell \ell+2, 00}^{BB}\equiv \frac{T_0^2 \ell (\ell+1)}{2\pi}\, C_{\ell \ell+2, 00}^{BB}\, .
\ee

Other off-diagonal elements of the correlation function are nonvanishing, like $\ell'=\ell+4$, $\ell'=\ell+6$, etc., and for any $m$ or $m'$. 

Finally, in an anisotropic background cross-correlations between B-mode polarization and the temperature or E-mode polarization will also be non-zero when $\ell + \ell'$ is odd. Further, only the primordial cross-correlations ${{\cal P}}_{\pm 2\mp 2}(\vec k)$, ${{\cal P}}_{0\pm 2}(\vec k)$ and ${{\cal P}}_{\pm 2 0}(\vec k)$ contribute, with all of these terms involving tensorial modes. As a result, the amplitude of these cross-correlations is model-dependent, with models that generate significant tensor modes during contraction (like the matter bounce) expected to lead to larger cross-correlations than what is presented here for the case where the tensor modes are vanishing a few $e$-folds before the bounce.

\begin{figure}[th]
{\centering
\includegraphics[width = 0.49\textwidth]{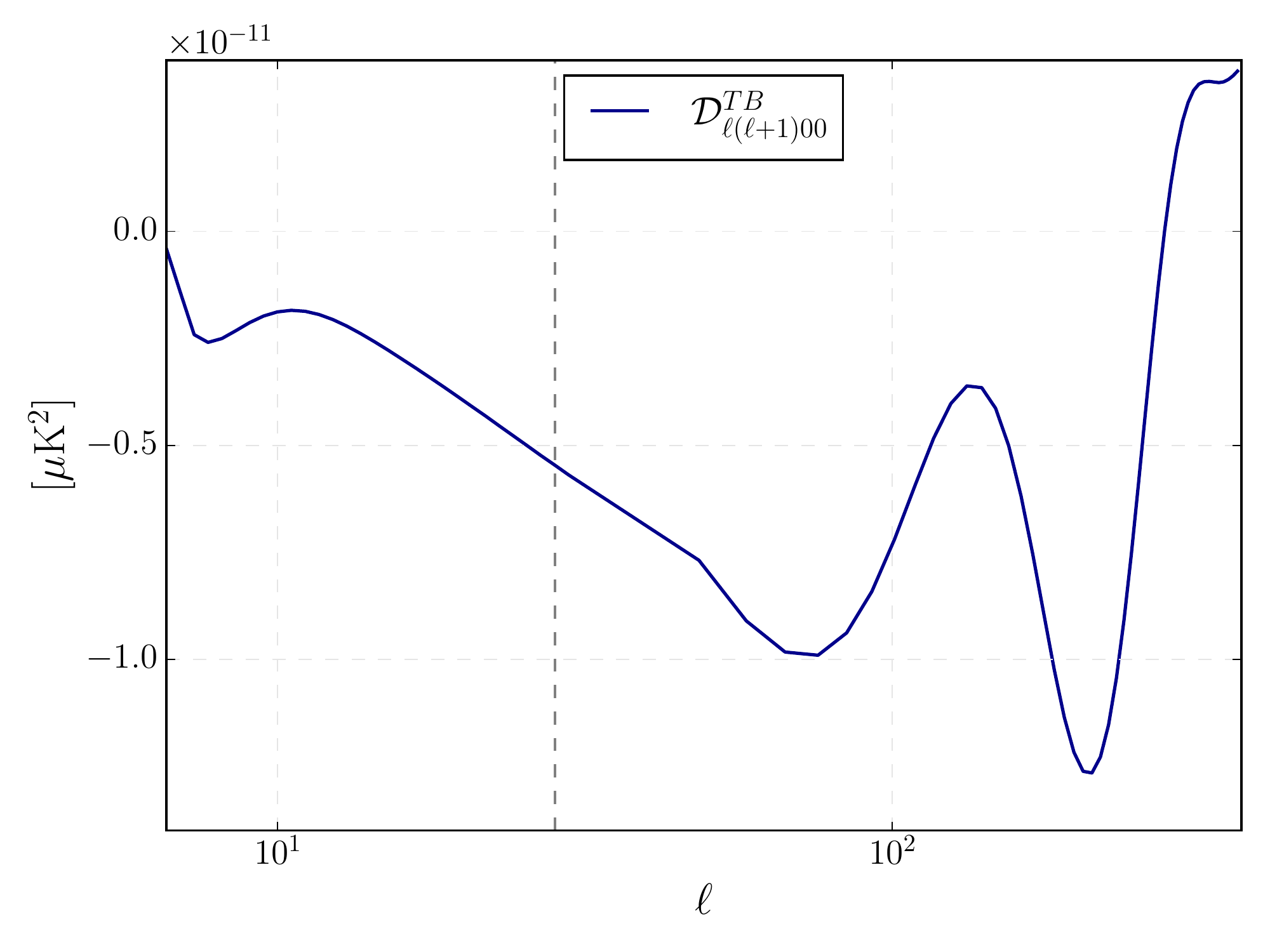}
\includegraphics[width = 0.49\textwidth]{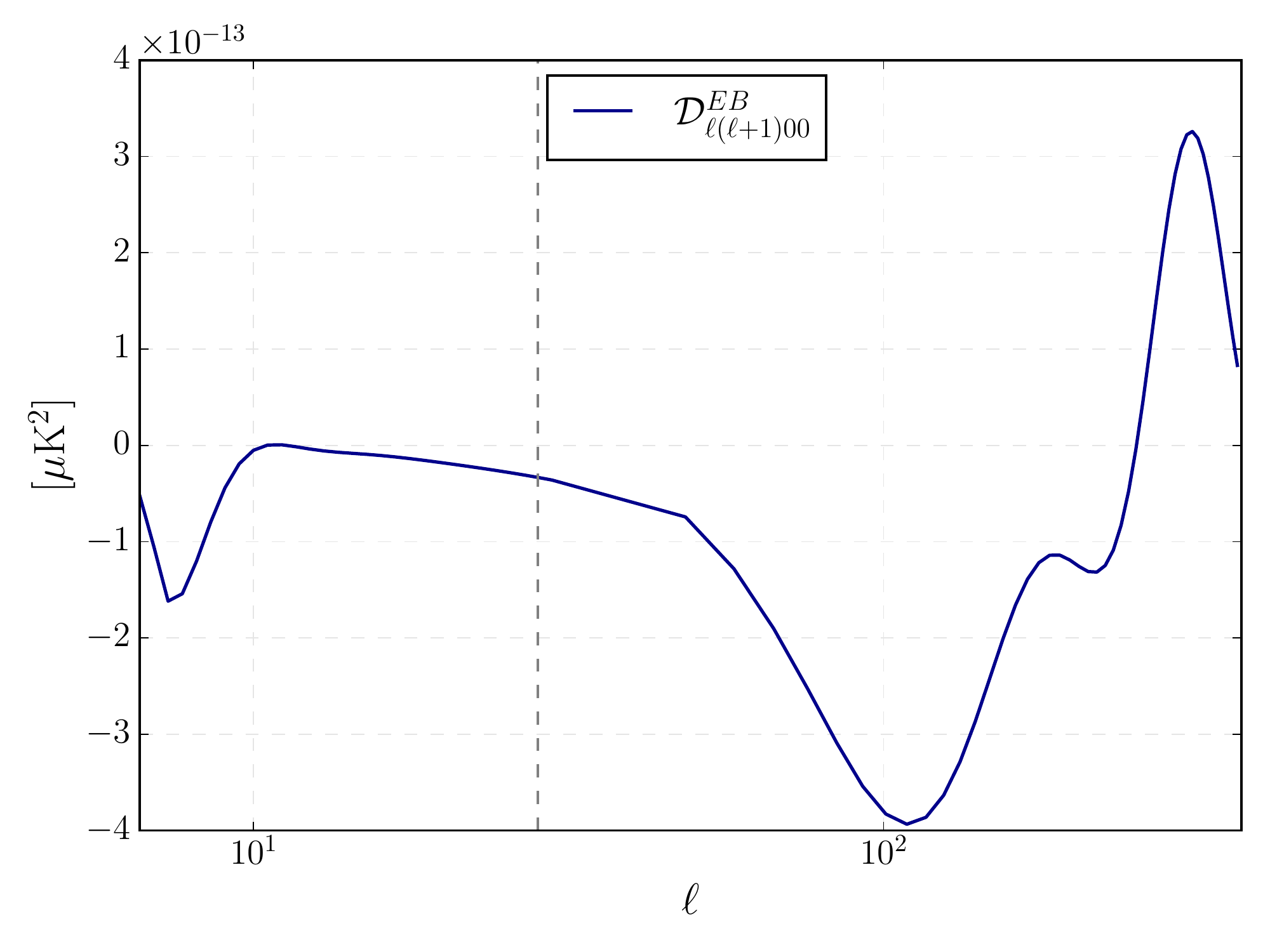}
}
\caption{T-B and E-B angular cross-correlation functions for perturbations on a radiation-dominated Bianchi~I background with the initial conditions described in Sec.~\ref{upp}. The vertical dashed line denotes the transition from a linear scale for $\ell$ (on its left) to a logarithmic scale (on its right). \\
\textit{Left panel:} The T-B angular correlation function $D_{\ell\ell+1,00}^{TB}$. This correlation function would be identically zero in absence of primordial anisotropies. \\
\textit{Right panel:} The E-B angular correlation function $D_{\ell \ell+1, 00}^{EB}$. This correlation function would be identically zero in absence of primordial anisotropies.}
\label{fig:TB}
\end{figure}

In Fig.~\ref{fig:TB}, we present two examples of angular cross-correlation functions, in the left panel is plotted
\be
D_{\ell\ell+1,00}^{TB}=\frac{T_0^2 \ell (\ell+1)}{2\pi}\, C_{\ell \ell+1, 0 0}^{TB}\, ,
\ee
and the right panel shows
\be
D_{\ell\ell+1,00}^{EB}=\frac{T_0^2 \ell (\ell+1)}{2\pi}\, C_{\ell \ell+1, 0 0}^{EB}\, ,
\ee
where we have chosen $\ell' = \ell + 1$ and $m=m'=0$ as a representative case for both of the cross-correlation functions. Other correlation functions with $\ell' = \ell + 3$, $\ell' = \ell + 5$, etc., and other values of $m,m'$ are non-zero. Moreover, all T-B and E-B cross-correlations would be identically zero if the power spectra of perturbations were isotropic.

In summary, primordial anisotropies leave distinct signatures in the CMB, in particular non-zero off-diagonal T-T, T-E, E-E and B-B correlation functions (with $\ell'=\ell + 2n$), as well as off-diagonal T-B and E-B cross-correlation functions (with $\ell'=\ell + 2n + 1$). Given observational constraints on the amplitude of B-B correlations and on anisotropic features in the CMB, the largest effect is the $\ell'=\ell+2$ off-diagonal T-T correlation function, shown in the right panel of Fig.~\ref{fig:DTT}, which would be identically zero for an isotropic background. This effect is largely captured by the quadrupole moment $g_2$, which we focused on in Secs.~\ref{dyn} and~\ref{upp}.

\section{Conclusion and Discussion}

Since anisotropies grow in a contracting universe, some degree of anisotropy is expected to be unavoidable in cosmological scenarios with a cosmic bounce. In this paper, we have performed a detailed study of how these anisotropies have an impact on cosmological perturbations, and on predictions for the cosmic microwave background for bouncing alternatives to inflation. The general features we find are generated in the contracting phase before the bounce, and are therefore to a large degree independent of the physics of the bounce.

Previously, constraints on the allowed growth of anisotropies in a contracting universe have been derived by requiring that the ratio $\sigma^2/(16 \pi\rho)$ must always remain smaller than one. We have found that CMB observations provide considerably stronger constraints on anisotropies than simply considering this ratio. One reason for this is that, while anisotropies in the background geometry rapidly decrease in an expanding universe, perturbations retain a memory of the primordial anisotropies, in the sense that the amplitude of anisotropic features in the perturbations does not decrease as the universe expands. (The expansion will redshift the wavelengths of the Fourier modes of the perturbations, but this does not affect the amplitude of a scale-invariant quantity.) Also, even small anisotropies in the background space-time can, over a sufficiently long time, generate observationally significant effects in the perturbations---it is not only the relative amplitude of anisotropies that is relevant, but the timescale is important as well. 

The leading order effect is a scale-invariant quadrupolar modulation $g_2$ of the primordial power spectrum (there is no dipole due to parity invariance). Due to the predicted scale-invariance of this effect, CMB observations provide strong constraints on anisotropies. This is in contrast to inflationary models, where primordial anisotropies can also generate a quadrupole moment in the CMB, but for inflationary models the quadrupole moment is predicted to decay rapidly for large $k$. For this reason, observational constraints on primordial anisotropies are stronger for bouncing alternatives to inflation. Also, if a quadrupole moment is observed, its scale-dependence can be used to distinguish between inflation and bouncing alternatives to inflation.

The constraints on primordial anisotropies, based on the quadrupole moment $g_2$, depend on the dominant matter field, especially during the bounce. As expected, these constraints are in some ways weaker for ekpyrotic models, but the bounds remain surprisingly strong even for ekpyrosis. This is because the perturbations are sensitive to the directional anisotropies $\sigma_i$ (which grow in a contracting universe, independently of the matter content), and the amplitude of anisotropic features in the perturbations remains constant in an expanding universe after the bounce (in stark contrast to the shear $\sigma^2$ which rapidly decays).

Also, the perturbations depend on the anisotropies in a rather complicated way, in particular depending on their history and not only on the maximal value of ratios like $\sigma^2/16 \pi \rho$. This makes it difficult to obtain universal constraints on anisotropies that could be extended to other models---as far as we can tell, to obtain constraints for any particular cosmological scenario it is necessary to solve the dynamics for the perturbations from scratch in that specific case.

We note that the simulations indicate that $g_2$ depends on $\sigma^2$, and that $g_2$ is independent of how its amplitude is distributed among the $\sigma_i$ as determined by $\Psi$; however, the value of $\Psi$ does have an effect on some other observables like the angular correlation functions $C^{XY}_{ll',mm'}$.

Another effect of anisotropies is to introduce an effective coupling between the scalar and the two tensor modes. Given observational constraints on the possible amplitude of tensor modes, and on the strength of primordial anisotropies, the effects due to this coupling are subleading compared to the quadrupole moment $g_2$. One effect of this coupling is to excite tensor modes, although the tensor-to-scalar ratio remains small for observationally allowed primordial anisotropies (assuming the tensor modes are initially vanishing, as is predicted by ekpyrotic models).

Anisotropies can also generate non-zero angular correlation functions in the CMB that would otherwise vanish in an isotropic background. For example, off-diagonal correlation functions $C_{\ell \ell'}^{TT}$ for $\ell'=\ell+2$ are non-zero (while the more familiar diagonal $\ell=\ell'$ correlation functions are not significantly impacted). Another smoking gun signal are T-B and E-B cross-correlations for odd $\ell+\ell'$; they must be small due to constraints on primordial gravitational waves and on primordial anisotropies, but nonetheless these are otherwise unexpected signals that can be generated in an anisotropic background.

In summary, by studying cosmological perturbations it is possible to understand the concrete imprints that anisotropies would leave in the CMB for bouncing alternatives to inflation, and obtain much stronger observational constraints on anisotropies than would otherwise be possible. Our analysis in this paper has been deliberately broad; it would be interesting to use these tools to constrain particular cosmological models of interest, whether specific realizations of ekpyrosis, the matter bounce, or other bouncing alternatives to inflation.

\acknowledgments

We thank J\'er\^ome Quintin for helpful comments on an earlier version of this paper.
I.A.~is supported by the NSF grant PHY-2110273, and by the Hearne Institute for Theoretical Physics.
J.O.~is supported by the Spanish Government through the projects PID2020-118159GB-C43, PID2019-105943GB-I00 (with FEDER contribution), and also by the ``Operative Program FEDER2014-2020 Junta de Andaluc\'ia-Consejer\'ia de Econom\'ia y Conocimiento'' under project E-FQM-262-UGR18 by Universidad de Granada.
E.W.-E.~is supported by the Natural Sciences and Engineering Research Council of Canada, and by the UNB Fritz Grein Research Award.

\appendix

\section{Cosmological perturbations on Bianchi~I space-times}
\label{Bianchi}

In this Appendix we review the dynamics of the Bianchi~I space-time in general relativity, and then discuss cosmological perturbation theory on a Bianchi~I background.

\subsection{Background Bianchi~I geometry}
\label{BI}

We focus on Bianchi~I space-times for which there exists a coordinate system in which the line element takes the diagonal form 
\be
\dd s^2 = -\dd t^2 + a_1^2(t) \, \dd x_1^2 + a_2^2(t) \, \dd x_2^2 + a_3^2(t) \, \dd x_3^2 \, , 
\ee
where $a_1$, $a_2$ and $a_3$ are independent directional scale factors, and for the matter sector we consider a scalar field $\phi$ driven by a potential $V(\phi)$.

The equations of motion for the four degrees of freedom $a_i(t)$ and $\phi(t)$ can be simplified by separating the geometric variables into the mean scale factor $a = (a_1 a_2 a_3)^{1/3}$ and purely anisotropic degrees of freedom given by the shears
\be \label{defsi}
\sigma_i \equiv H_i - H \, ,
\ee
where $H=\frac{d\ln a}{dt}$ and $H_i=\frac{d\ln a_i}{dt}$ are the mean and directional Hubble rates, respectively. It is also often convenient to introduce the total shear squared $\sigma^2 = \sigma^2_1 + \sigma^2_2 + \sigma^2_3$. Note that the definition of $\sigma_i$ implies that $\sigma_1 + \sigma_2 + \sigma_3 = 0$, so only two of these variables are independent.

The equations of motion for these degrees of freedom can be derived from the Einstein equations. The off-diagonal components of these equations imply that $\dot \sigma_i(t) = -3\, H\, \sigma_i$, and therefore $a^3\, \sigma_i$ are constants of the motion. The two independent constants of the motion can be conveniently encoded in two constant non-negative real numbers, $\Sigma$ and $\Psi$, in the following way (see, e.g., \cite{Agullo:2020uii})
\bea \label{shearsev}
\sigma^2(t)&=\Sigma^2 / a^6(t) \, , \qquad \qquad \qquad \qquad \qquad
\sigma_1(t)&=\sigma(t) \, \sqrt{\frac{2}{3}}\, \sin \Psi\, , \\
\sigma_2(t)&=\sigma(t) \, \sqrt{\frac{2}{3}}\, \sin \left(\Psi+\frac{2\pi}{3}\right) \, , \qquad \qquad
\sigma_3(t)&=\sigma(t)\, \sqrt{\frac{2}{3}}\, \sin \left(\Psi+\frac{4\pi}{3}\right)\, \nonumber .
\eea
As can be seen from these expressions, $\Sigma$ controls the amplitude of the anisotropies, while $\Psi$ determines how the anisotropies are distributed among the principal directions.

The equations of motion for the isotropic degrees of freedom $a(t)$ and $\phi(t)$ come from the diagonal components of Einstein's equations, which give two dynamical equations
\be \label{aver}
\frac{\ddot a}{a}=-\frac{2\pi}{3}\, (\rho+3\, p)-\frac{\sigma^2}{3} \, ; \hspace{1cm} \ddot \phi+3\f{\dot a}{a}\, \dot\phi+\frac{\d V(\phi)}{\d\phi}=0 \, ,
\ee
and one constraint 
\be \label{cons}
H^2=\frac{8\pi}{3}\, \rho +\frac{\sigma^2}{6}\, ,
\ee
where $\rho=\frac{1}{2}\dot \phi^2+V(\phi)$ and $p=\frac{1}{2}\dot \phi^2-V(\phi)$.

To solve these equations, we need to provide $\Sigma$ and $\Psi$ as well as initial data for $a$, $\dot a$, $\phi$, and $\dot \phi$ that are compatible with the constraint. The initial value for the scale factor $a$ can be chosen freely, since this is not a physical observable and the value of $a$ can be arbitrarily changed by a rescaling of the coordinates $x_i$. Because of this, the physical input for the initial conditions (for given $\Sigma$ and $\Psi$) can be determined from $\rho$, $w$, and the sign of $\dot a$, by using \eqref{cons}.

In summary, a unique solution for the background geometry and the scalar field can be obtained by specifying the initial values for $\rho$, $w$, and the sign of $H$ together with $\Sigma$ and $\Psi$ for the anisotropies. From this, $\sigma_i(t)$ can be calculated from \eqref{shearsev}, and solving the differential equations \eqref{aver} gives $a(t)$ and $\phi(t)$. The constraint \eqref{cons} is guaranteed to be preserved dynamically by the consistency of Einstein equations, while the evolution of the directional scale factors can be easily obtained by solving the differential equation $\frac{\dot a_i}{a_i} = H_i = \sigma_i(t) + H(t)$.

\subsection{Cosmological perturbations}
\label{app.perts}

As stated in Sec.~\ref{subsec:perts}, linear perturbations on a Bianchi~I background spacetime can be written in terms of three gauge-invariant fields $\Gamma_{\mu}(\vec x)$, with $\mu=0,1,2$, that reduce to the familiar scalar and tensor perturbations in the isotropic limit---more precisely, $\Gamma_0$ becomes $\sqrt{32\pi }\, \frac{z}{a}\, \mathcal{R}(\vec{k})$, where $\mathcal{R}(\vec{k})$ is the familiar comoving curvature perturbation and $z=a\, \frac{\dot \phi}{H}$, while $\Gamma_1$ and $\Gamma_2$ reduce to the $+,\times$ polarizations of tensor modes, respectively.

The equations for the $\Gamma_\mu$ on a Bianchi~I background are \cite{Pereira:2007yy, Agullo:2020uii}
\be
\ddot \Gamma_{\mu}(\vec k)+3\, H\, \dot \Gamma_{\mu}(\vec k)+\frac{k^2}{a^2}\, \Gamma_{\mu}(\vec k)+\frac{1}{a^2}\, \sum_{\mu'=0}^2\, {\cal U}_{\mu\mu'}(\hat k, t) \, \Gamma_{\mu'}(\vec k)=0\, ,
\ee
where dots denote derivatives with respect to $t$, and
\be
k^2(t) \equiv a^2(t)\, \left( \frac{k_1^2}{a_1^2(t)} + \frac{k_2^2}{a_2^2(t)} + \frac{k_3^2}{a_3^2(t)} \right).
\ee
The effective potentials ${\cal U}_{\mu\mu'}(\hat k, t)$ are
\bea
\, {\cal U}_{00}&=&a^2\, V_{\phi\phi}\, - \frac{2 \kappa \, \pp^2 {\cal F}_2}{a^3} + 2 \kappa \, {\cal F}_1 \left(-\frac{\kappa\,\pp^2\,p_a}{3a^{5}}\, + \,2 \,V_{\phi}\, \pp\right), \\ \nonumber
{\cal U}_{01}&=&{\cal U}_{10} \,= \,\frac{2\sqrt{\kappa}}{a^2}\left(-a^2\,\pp\, \sigma_{(5)} \, {\cal F}_2 + a^5V_{\phi} \, \sigma_{(5)}\, {\cal F}_1 - a^2\,\pp \, {\cal G}_{5}\, {\cal F}_1\, +\, \frac{\kappa}{6}\,\pp\,p_a\,\sigma_{(5)}\,{\cal F}_1\right)\,, \\ \nonumber
{\cal U}_{02}&=&{\cal U}_{20}\,= \,\frac{2\sqrt{\kappa}}{a^2}\left(-a^2\,\pp\, \sigma_{(6)} \, {\cal F}_2 + a^5\,V_{\phi} \, \sigma_{(6)}\, {\cal F}_1 - a^2\,\pp \, {\cal G}_{6}\, {\cal F}_1\, +\, \frac{\kappa}{6}\,\pp\,p_a\,\sigma_{(6)}\,{\cal F}_1\right)\,, \\ \nonumber
{\cal U}_{12}&=&{\cal U}_{21}\,=\, 2\,\sigma_{(5)}\,\sigma_{(6)}\,\left( a^2 -\, a^3\,{\cal F}_2\,+\, \frac{2}{3}\,\kappa\,a\,p_a\,{\cal F}_1\right) - \left( \, 2\,a^3\,\sigma_{(6)}\,{\cal G}_{5}\, +\, 2\,a^3\,\sigma_{(5)}\,{\cal G}_{6}\right)\,{\cal F}_1\, , \\ \nonumber
{\cal U}_{11}\, &=&\,- 2 \,a^2\, \sigma_{(6)}^2\, +\,\frac{\kappa p_a \,\sigma_{(2)}}{\sqrt{6}}\,-\,a^2\,\sqrt{\f{2}{3}}{\cal G}_2\, +\,\frac{4}{3}\,\kappa\,a\,p_a\,\sigma_{(5)}^2\,{\cal F}_1\,-\,4 \,a^3\, \sigma_{(5)}\,{\cal F}_1 \, {\cal G}_5\,-\,2\,a^3\, \sigma_{(5)}^2\,{\cal F}_2\, , \\ \nonumber
{\cal U}_{22}\, &=&\,- 2 \,a^2\, \sigma_{(5)}^2\, +\,\frac{\kappa p_a \,\sigma_{(2)}}{\sqrt{6}}\,-\,a^2\,\sqrt{\f{2}{3}}{\cal G}_2\, +\,\frac{4}{3}\,\kappa\,a\,p_a\,\sigma_{(6)}^2\,{\cal F}_1\,-\,4 \,a^3 \,\sigma_{(6)}\, {\cal F}_1\, {\cal G}_6\,-\,2\,a^3\, \sigma_{(6)}^2\, {\cal F}_2\,,
\eea
with $V_{\phi} \equiv dV/d\phi$, $V_{\phi\phi} \equiv d^2V/d\phi^2$, and
\begin{eqnarray}
{\cal F}_1\, &=&\, \frac{-\frac{\kappa p_a}{2a^3}\, +\,\sqrt{\frac{3}{2}} \,\frac{\sigma_{(2)}}{a}}{2\kappa\rho\,+\,\sigma_{(3)}^2\,+\, \sigma_{(4)}^2+\, \sigma_{(5)}^2\,+\, \sigma_{(6)}^2},\\ \nonumber
{\cal F}_2\, &=&\frac{\frac{3\kappa \, V}{a}\, - \,\frac{\kappa^2 p_a^2}{3a^5}\,+\,\frac{\kappa p_a\sigma_{(2)}}{2\sqrt{6}a^3}\, +\, \sqrt{\frac{3}{2}}\frac{{\cal G}_{2}}{a}\,-\,{\cal F}_1\left[\frac{\kappa^2 \pp^2p_a}{a^8}\,+\,2\,\sigma_{(3)}\, {\cal G}_3\,+\,2\,\sigma_{(4)}\, {\cal G}_4+\, 2\,\sigma_{(5)}\, {\cal G}_5\,+\, 2\,\sigma_{(6)}\, {\cal G}_6)\right]}{2\kappa\rho\,+\,\sigma_{(3)}^2\,+\, \sigma_{(4)}^2+\, \sigma_{(5)}^2\,+\, \sigma_{(6)}^2},\\ \nonumber
{\cal G}_2 &=& \frac{\kappa\, p_a\sigma_{(2)}}{2\,a^2}\, -\, \sqrt{\frac{3}{2}}\left(\sigma_{(3)}^2 \,+\, \sigma_{(4)}^2\right),\\ \nonumber
{\cal G}_3 &=& \frac{\kappa\, p_a\, \sigma_{(3)}}{2\,a^2}\, +\, \frac{1}{\sqrt{2}} \left(\sqrt{3}\sigma_{(2)} \sigma_{(3)} - \sigma_{(3)} \sigma_{(5)} - \sigma_{(4)} \sigma_{(6)}\right),\\ \nonumber
{\cal G}_4 &=& \frac{\kappa\, p_a\sigma_{(4)}}{2\,a^2} + \frac{1}{\sqrt{2}} \left(\sqrt{3}\sigma_{(2)} \sigma_{(4)} + \sigma_{(4)} \sigma_{(5)} - \sigma_{(3)} \sigma_{(6)}\right)\,, \\ \nonumber
{\cal G}_5 &=& \frac{\kappa\, p_a\sigma_{(5)}}{2\,a^2}\, +\, \frac{1}{\sqrt{2}}(\sigma_{(3)}^2 - \sigma_{(4)}^2),\\ \nonumber
{\cal G}_6 &=& \frac{\kappa\, p_a\sigma_{(6)}}{2\,a^2} + \sqrt{2}\,\sigma_{(3)}\sigma_{(4)}.
\end{eqnarray}
In these expressions, $\pp=a^3\, \dot \phi$, $p_a=-\frac{6}{\kappa}\, a^2 \, H$ and $\kappa = 8 \pi \, G$. The quantities $\sigma_{(n)}$ can be calculated from the shear tensor $\sigma_{ij}={\rm diag}(a_1^2\, \sigma_1,a_2^2\, \sigma_2,a_3^2\, \sigma_3)$; specifically, $\sigma_{(n)} := \sigma_{ij} \, A_{(n)}^{ij}$ (with $n=1,\ldots,6$) are the projections of the shear tensor $\sigma_{ij}$ on the matrices $A_{(n)}^{ij}$ commonly used to implement the scalar-vector-tensor decomposition for cosmological perturbations on FLRW backgrounds in Fourier space:
\begin{align}
{A}^{{(1)}}_{ij}\, &=\, \f{h_{ij}}{\sqrt{3}}, \hspace{0.5in} & {A}^{(4)}_{ij}\, &=\,\f{1}{\sqrt{2}}\, \l(\, \hat{ k}_i\, \h y_j\, +\, \hat{ k}_j\, \h y_i \,\r),\nonumber\\
{A}^{(2)}_{ij}\, &=\,\sqrt{\f{3}{2}}\,\l(\hat{ k}_i\,\hat{ k}_j - \f{h_{ij}}{3}\r), \hspace{.5in} &{A}^{(5)}_{ij}\,& = \, \f{1}{\sqrt{2}}\, \l(\, \hat x_i\, \h x_j\, -\, \hat y_i\, \h y_j \,\r), \nonumber\\
{A}^{(3)}_{ij}\, &=\, \f{1}{\sqrt{2}}\, \l(\, \hat{ k}_i\, \h x_j\, +\, \hat{ k}_j\, \h x_i \,\r),
\hspace{.5in} &{A}^{(6)}_{ij}\, &=\,\f{1}{\sqrt{2}}\, \l(\, \hat x_i\, \h y_j\, +\, \hat x_j\, \h y_i \,\r). \label{matrixbases}
\end{align}
The $i,j$ are spatial indices that are raised and lowered with the spatial part of the Bianchi~I metric, $h_{ij} = {\rm diag}(a_1^2, a_2^2, a_3^2)$. In addition, $\hat{k}$ is the unit vector in the direction of $\vec{k}$, normalized with respect to $h_{ij}$. $\h x$ and $\h y$ are two unit vectors which, together with $\hat k$, form (a time-dependent) orthonormal triad, with an orientation satisfying $\hat x \times \hat y = \hat k$.

Note that $\sigma_{(n)}$ depends on $\hat k$, since the matrices ${A}^{(n)}_{ij}$ do, and it is because of this dependence that the potentials ${\cal U}_{\mu\mu'}$ depend on $\hat k$. Note however that the potentials ${\cal U}_{\mu\mu'}$ are independent of the modulus $k$ of the wavevector. Note also that $\sigma_{(n)}$ should not be identified as the Fourier components of the tensor $\sigma_{ij}$, since $\sigma_{ij}$ is position independent and its Fourier transform is trivial. The quantities $\sigma_{(n)}(\hat k)$ are just a convenient way of decomposing the shear tensor, adapted to the equations of motion for the perturbations $\Gamma_\mu$.

For further details regarding cosmological perturbation theory on an anisotropic Bianchi~I background, see Refs.~\cite{Pereira:2007yy, Agullo:2020uii}.

\section{Loop quantum cosmology}
\label{LQC}

This appendix provides a brief summary on the effective equations of loop quantum cosmology (LQC) for Bianchi~I space-times. Although we use LQC to generate the bounce in the simulations, the focus of this work is on the contracting pre-bounce phase. For this reason, we will only briefly summarize the equations that we solve, in order to make our article self-consistent. For further details about the physics of LQC and the equations we show below, see \cite{Ashtekar:2009vc, Ashtekar:2011ni, Agullo:2016tjh}.

In LQC, the spacetime geometry is quantum, and it is describe by a wave function $\Psi$ belonging to a suitable Hilbert space. The metric tensor, curvature, etc., are operators acting on these states. To make contact with semi-classical physics, it is of interest to consider quantum states $\Psi$ that are sharply peaked on a classical geometry at late times, when general relativity becomes an excellent approximation. It turns out that for this class of states, the expectation values of observables follow trajectories given by the so-called effective dynamics that can be derived from an effective Hamiltonian that includes some $\hbar$ corrections \cite{Taveras:2008ke, Diener:2014mia}. While studies of the validity of the LQC effective equations as an approximation to the full quantum dynamics of sharply-peaked states have focused on isotropic FLRW space-times, LQC effective equations are also expected to hold for anisotropic space-times, since their validity is due to the fact that the phase space variables correspond to large-scale degrees of freedom \cite{Rovelli:2013zaa, Bojowald:2015fla}.

Since these effective equations of motion in LQC are derived in a Hamiltonian formalism, it is helpful to briefly summarize the Hamiltonian framework in standard general relativity, applied to Bianchi~I space-times. The phase space for Bianchi cosmologies (coupled to a scalar field) is coordinatized by four pairs of canonically conjugate variables, namely $a_1,a_2,a_3,\phi$ and their conjugate momenta $\pi_{a_1},\pi_{a_2},\pi_{a_3},\pp$, for which the non-vanishing Poisson brackets are
\be \label{bpb}
\{\phi, \pp\}=\frac{1}{\mathcal{V}_0}\, , \hspace{0.5cm} \{a_i, \pi_{a_j} \}=\frac{1}{\mathcal{V}_0} \, \delta_{ij}\, ,
\ee
where $\mathcal{V}_0$ is the (coordinate) volume of space%
\footnote{Because of the homogeneity of space, the integrals involved in the definition of the Hamiltonian and the symplectic form diverge if the spatial slice is non-compact. But this divergence is spurious, and it can be regularized by restricting integrals to a box of arbitrarily large but finite coordinate volume $\mathcal{V}_0$. It is convenient to choose the sides of the box to be aligned with the three principal axes $x_i$, and of length $L_i$, so $\mathcal{V}_0 = L_1 L_2 L_3$. This volume is an infrared regulator, which can be taken to infinity at the end of the calculation.}.
The dynamics of the Bianchi~I geometry and the scalar field can be obtained from Hamilton's equations, with the Hamiltonian given by the constraint \eqref{cons}
\bea \label{Fcons}
\mathcal{H}_{\rm BI} &= \f{\mathcal{V}_0}{2\sqrt{h}} \biggl[ &
\kappa \l( \f{a_1^2 \pi_{a_1}^2}{2} + \f{a_2^2 \pi_{a_2}^2}{2} + \f{a_3^2 \pi_{a_3}^2}{2} 
- a_1 \pi_{a_1} a_2 \pi_{a_2} - a_2 \pi_{a_2} a_3 \pi_{a_3} - a_3 \pi_{a_3} a_1 \pi_{a_1} \r)
\nonumber\\&& ~
+ \, p_{\phi}^2 + 2 h \, V(\phi) \biggr],
\eea
where $h=(a_1a_2a_3)^2$ is the determinant of spatial metric $h_{ij}$, and $\kappa=8\pi G$. Hamilton's equations
\bea \label{eoma}
\dot a_i &= \{a_i, \mathcal{H}_{\rm BI} \} , \qquad \qquad \dot \pi_{a_i}&=\{\pi_{a_i}, \mathcal{H}_{\rm BI}\} \, , \\ \nonumber 
\label{eomphi} \dot \phi &= \{\phi, \mathcal{H}_{\rm BI}\} \, , \qquad \qquad \dot{\pp} &= \{\pp, \mathcal{H}_{\rm BI} \} \, ,
\eea
are equivalent to Einstein's equations for Bianchi~I space-times, described in App.~\ref{BI}. 
 
The equations of LQC are written in terms of a different set of variables, due to the use of a connection rather than a metric tensor to describe the gravitational field---these are the Ashtekar-Barbero variables \cite{Ashtekar:1986yd, BarberoG:1994eia}. For Bianchi~I geometries, these variables reduce to three pairs of canonically conjugate variables $c_i$ and $p_i$, with $i=1,2,3$, for the geometric degrees of freedom. The non-vanishing Poisson brackets are
\be
\{c_i,p_j\}= \f{\kappa \gamma}{{\mathcal{V}_0}} \, \delta_{ij}, \quad \left\{\phi, p_{\phi}\right\}=\frac{1}{\mathcal{V}_{0}}\, ,
\ee
where $\gamma$ is a constant known as the Barbero-Immirzi parameter; $\gamma$ has no effect on classical physics, but its value does have an impact on some quantum gravity effects.

The relation between $c_i$ and $p_i$ and the metric variables $a_i$ and $\pi_{a_i}$ is
\be
p_i=\frac{\mathcal{V}_0}{L_i}\, \frac{a^3}{a_i}\, , \ \ \ \ {c_i=-\kappa \, \gamma \,L_i\, a^{-3} a_i \, \Big(a_i\, \pi_{a_i}-\frac{1}{2}\sum_j a_j\pi_{a_j} \Big)}\, , \ \ \ \ ({\rm no\ sum \ in \ }i)\,.
\ee
The classical Hamiltonian constraint (\ref{Fcons}), when written in terms of $c_i$ and $p_i$, takes the form
\be\label{calssHAv}
\mathcal{H}_{\rm BI}\, =\left[\f{-1}{\kappa\, \gamma^2 \,v}\biggl( 
c_1 c_2p_1 p_2\, +\,c_1 c_3p_1p_3\,+\,c_2 c_3p_2p_3\,\biggr)\, +\, \f{{\cal V}^2_0p_{\phi}^2}{2v}\, + {v\, V(\phi)}\, \right],
\ee
where $v=\sqrt{p_1p_2p_3}$. A physical solution to the equations of motion 
\begin{align} \label{lqceqs}
\dot{c}_i &=\left\{c_i, \mathcal{H}_{\mathrm{BI}}\right\} , \quad \dot{p}_i=\left\{p_i, \mathcal{H}_{\mathrm{BI}}\right\} ,\\ \dot{\phi} &=\left\{\phi, \mathcal{H}_{\mathrm{BI}}\right\} , \quad \dot p_{\phi}=\left\{p_{\phi}, \mathcal{H}_{\mathrm{BI}}\right\}, 
\end{align}
is uniquely singled out by specifying $\phi(t_0)$, $c_i(t_0)$ for $i=1,2,3$, and the sign of $\pp(t_0)$ at any instant $t_0$. (In exactly the same way as for the directional scale factors $a_i$, the concrete numerical values of $p_i(t_0)$ are not observable as they change under rescalings of the coordinates $x_i$, and different choices for $p_i(t_0)$ produce physically equivalent solutions). So far this is completely equivalent to general relativity, merely written in other variables.

The leading order quantum effects in LQC can be incorporated through the replacement \cite{Ashtekar:2009vc}
\be
c_i \to \f{\sin(\bar\mu_i\,c_i)}{\bar\mu_i},
\ee
where $\bar\mu_1 \equiv \sqrt{\Delta} \,\sqrt{p_1 / p_2\,p_3}$, and similarly for $\bar \mu_2$ and $\bar\mu_3$, where $\Delta = 4 \sqrt{3} \pi \gamma G \hbar$ is the smallest non-zero eigenvalue of the area operator in LQG. These trigonometric functions capture in a precise way the dominant quantum effects in LQC. As a result, the LQC effective Hamiltonian is
\begin{eqnarray} \label{lqc-ham}
\mathcal{H}^{\rm LQC}_{\rm BI} = N \bigg[ & {\f{-1}{\kappa\,\gamma^2\,v}} \biggl( &
\f{\sin(\bar\mu_1\,c_1)}{\bar\mu_1}\f{\sin(\bar\mu_2\,c_2)}{\bar\mu_2}p_1p_2\, +\f{\sin(\bar\mu_1\,c_3)}{\bar\mu_1} \f{\sin(\bar\mu_3\,c_3)}{\bar\mu_3}p_1p_3\nonumber\\ && ~
+\f{\sin(\bar\mu_2\,c_2)}{\bar\mu_2} \f{\sin(\bar\mu_3\,c_3)}{\bar\mu_3}p_2p_3\biggr)
+\f{\mathcal{V}_0^2p_{\phi}^2}{2 v}\, + v\, V(\phi)\bigg].
\end{eqnarray}
The solutions to Hamilton's equations \eqref{lqceqs} derived from this Hamiltonian are the effective equations of LQC for Bianchi~I space-times. Once a solution is obtained, the directional Hubble parameters are given by
\begin{align} \label{lqc-eoms}
H_1 &=\frac{1}{2 \gamma 
   \sqrt{\Delta} } \Big[\sin \left(\bar\mu_1c_1 -\bar\mu_2c_2 \right)+\sin \left(\bar\mu_1c_1 -\bar\mu_3c_3
   \right)+\sin \left(\bar\mu_2c_2 +\bar\mu_3c_3 \right)\Big], \nonumber \\
H_2 &= \frac{1}{2 \gamma 
   \sqrt{\Delta} } \Big[\sin \left(\bar\mu_2c_2-\bar\mu_1c_1 \right)+\sin \left(\bar\mu_2c_2 -\bar\mu_3c_3
   \right)+\sin \left(\bar\mu_1c_1 +\bar\mu_3c_3 \right)\Big], \\
H_3 &= \frac{1}{2 \gamma 
   \sqrt{\Delta}} \Big[\sin \left(\bar\mu_3c_3-\bar\mu_1c_1
   \right)+\sin \left(\bar\mu_3c_3-\bar\mu_2c_2 \right)+\sin \left(\bar\mu_1c_1 +\bar\mu_2c_2 \right)\Big], \nonumber
\end{align}
and these can be integrated to find the scale factors $a_i$. Also from these expressions, it is possible to calculate the mean Hubble rate $H=\frac{1}{3}\sum_{i=1}^3H_i$, as well as the components of the directional shears $\sigma_i=H_i-H$. The equations of motion for $\phi$ and its momentum $p_\phi$ are the same as in classical general relativity.

One consequence of the LQC corrections is that the shears $\sigma_i$ no longer scale exactly as $a^{-3}$, and therefore $\sigma^2 = \sum_i \sigma_i^2$ also no longer scales exactly as $\Sigma^2 / a^6$ with $\Sigma$ a constant. Rather, defining $\tilde \Sigma^2 = a^6 \sigma^2$, it can be checked that $\tilde \Sigma^2$ starts to vary during the bounce, but rapidly approaches the same constant value $\Sigma^2$ either side of the bounce in the classical limit \cite{Chiou:2007sp}; this result holds whether anisotropies are large or small.

Denoting the classical limit of $\tilde \Sigma^2$ as $\Sigma^2$, it is possible to derive a modified Friedman equation for the LQC effective dynamics of Bianchi~I space-times \cite{Chiou:2007sp}
\be \label{effBI}
H^2 = \f{8 \pi G}{3} \rho \left( 1 - \f{\rho}{\rho_b} \right) + \f{\Sigma^2}{a^6} - \f{3 \Sigma^2 \rho}{a^6 \rho_b} - \f{27 \Sigma^4}{8 \pi G a^{12} \rho_b},
\ee
up to higher order terms in $\Sigma^2/a^6$ that are negligible for the runs of interest here that have small anisotropies. Here $\rho_b$ is the critical energy density; it determines the curvature scale that the bounce occurs at (note that the presence of the last three terms in \eqref{effBI} implies that the energy density at the bounce does not necessarily coincide with $\rho_c$ if $\Sigma$ is different from zero).

In this paper, since $\Sigma^2 / a^6 \ll \rho$ in all our simulations (since otherwise the anisotropies would violate observational constraints), the cosmic bounce happens when the energy density is extremely close to $\rho_b$. Note that the departures from general relativity due to LQC cause a cosmic bounce to occur, but the LQC effects are entirely negligible away from the bounce. Specifically, the departures from general relativity are significant only for the very short time interval around the bounce, of the order of the time $\sim \rho_b^{-1/4}$ in Planck units.

For studies of the LQC bounce for Bianchi space-times with large anisotropies, see \cite{Gupt:2012vi, Wilson-Ewing:2017vju, Wilson-Ewing:2018lyx}.

\raggedright

\end{document}